\documentclass[12pt]{iopart}

\expandafter\let\csname equation*\endcsname\relax
\expandafter\let\csname endequation*\endcsname\relax
\usepackage{graphicx}
\usepackage{amsmath}
\usepackage{color}

\begin{document}

\title{Shortcut loading atoms into an optical lattice}

\author{Xiaoji Zhou$^1$, Shengjie Jin$^1$ and J. Schmiedmayer$^2$}

\address{$^1$ School of Electronics Engineering and Computer Science, Peking University, Beijing 100871, China}
\address{$^2$ Vienna Center for Quantum Science and Technology, Atominstitut, TU Wien, Stadionallee 2, 1020 Vienna, Austria}
\ead{\mailto{xjzhou@pku.edu.cn},\mailto{schmiedmayer@atomchip.org}}

\vspace{10pt}

\begin{abstract}
  We present an effective and fast (few microseconds) procedure for transferring ultra-cold atoms from the ground state in a harmonic trap into the desired bands  of an optical lattice. Our shortcut method is a designed pulse sequence where the time duration and the interval in each step are fully optimized in order to maximize robustness and fidelity of the final state with respect to the target state. The atoms can be prepared in a single band with even or odd parity, and superposition states of different bands can be prepared and manipulated. Furthermore, we extend this idea to the case of two-dimensional or three-dimensional optical lattices where the energies of excited states are degenerate. We experimentally demonstrate various examples and show very good agreement with the theoretical model. Efficient shortcut methods will find applications in the preparation of quantum systems, in quantum information processing, in precise measurement and as a starting point to investigate dynamics in excited bands.
\end{abstract}

\today

\vspace{2pc}

\newpage
\tableofcontents
\newpage

\section{Introduction}\label{Introduction}
Efficient preparation and manipulation of ultracold atomic gases in optical lattices (OL) have applications in many
fields, including quantum simulation of many-body systems, the
realization of quantum computation, quantum optics, and high-precision atomic clocks~\cite{BlochRev,Dervianko,Cronin,book}. There is a common concern
 how to quickly transfer the Bose-Einstein condensates (BEC) from the initial harmonic
trap into a desired band of an OL with high fidelity and robustness. For example, to load atoms into the
ground band in an OL, one chooses to ramp up the lattice depth adiabatically, the time scale usually
lasts up to tens of milliseconds.  To shorten the time of transfer, different techniques were proposed, sharing the concept of "shortcuts to adiabaticity"~\cite{reviewSTA13}. They promise to reach the same target state as the adiabatic process but within a very short time. One kind of shortcut is the continuous action method, including counter-diabatic driving, fast-forward protocols and inverse engineering. They are developed and exploited extensively in rapid manipulations of cold atoms, such as expansion/compression, rotation, transport and loading, etc.~\cite{Chen,Schaff-2,Oliver,Sofia,JorgeSciRep,Campo,Masuda,Strungari}. The other is optimal control~\cite{Rabitz,Jorge} or composite pulses like in nuclear magnetic resonance~\cite{Levitt}. These techniques have been used for atomic clocks, atomic interferometry and quantum computing~\cite{Schleier,Butts,Lee}. As for loading atoms into an OL, theoretical proposals such as adding a supplementary driving potential~\cite{Campo,Takahashi} are very attractive.

Recently ultracold gases in higher bands of OL attracted much attention. Many interesting many-body phenomena, e.g., supersolid quantum phases in cubic lattices~\cite{Scarola}, quantum stripe ordering in
triangular lattices~\cite{Wu}, orbital degeneracy~\cite{Lewenstein} can
appear with ultracold atoms in excited-band states. However, the most widely-used adiabatic approaches can not directly
transfer atoms into the excited bands. Several experimental techniques
have been developed including: (i) coherent manipulation of vibrational bands by stimulated Raman transitions~\cite{Muller}, (ii) using a moving lattice to load a BEC into an excited-band~\cite{Browaeys}%
, (iii) swapping population to selectively exciting the atoms
into the P-band~\cite{Wirth1} or F-band~\cite{M} of a bipartite
square OL. All these approaches required
to transfer atoms into the S-band firstly. Fast and high fidelity shortcut directly loading into the desired band is lacking.

In this paper, we demonstrate an effective method for transferring atoms from an harmonic trap into the desired band of an OL.
This shortcut stems from nonholonomic coherent control~\cite{Lloyd,Harel}, and is composed by standing-wave pulse sequences which are imposed on the system before the lattice is switched on. The time duration and interval in each step are optimized in order to reach the target state with a high fidelity and robustness. This process can be completed within several tens
of microseconds, reducing the loading time by up to three orders of magnitude as compared to adiabatic loading. It can be applied to load different excited bands and open up the possibility to study their dynamic behavior. Furthermore, we demonstrate the manipulation of the superposition of Bloch states and loading into two-dimensional (2D) and three-dimensional (3D) OL.  Our experimental results are in good agreement with the theoretical model.

The structure of this manuscript is organized as follows. In Sec.~\ref{Method}, we
introduce our idea of shortcut loading and optimization of pulse sequences to the S-band with zero quasi-momentum.  The demonstration of loading atoms into odd parity excited bands such as the D- or G-band, and even parity excited bands such as the P-band in OL are given in Sec.~\ref{Higher}.
In Sec.~\ref{Manipulation}, the shortcut loading atoms into S-band with non-zero quasi-momentum and into superpositions of band states are implemented.  The case of 2D or 3D OL with degenerate energies of the excited states are shown in Sec.~\ref{3D}. Finally, the main results are summarized in Sec.~\ref{Conclusions}.

\begin{figure}[t]
  \begin{center}
	\includegraphics[width=1\textwidth]{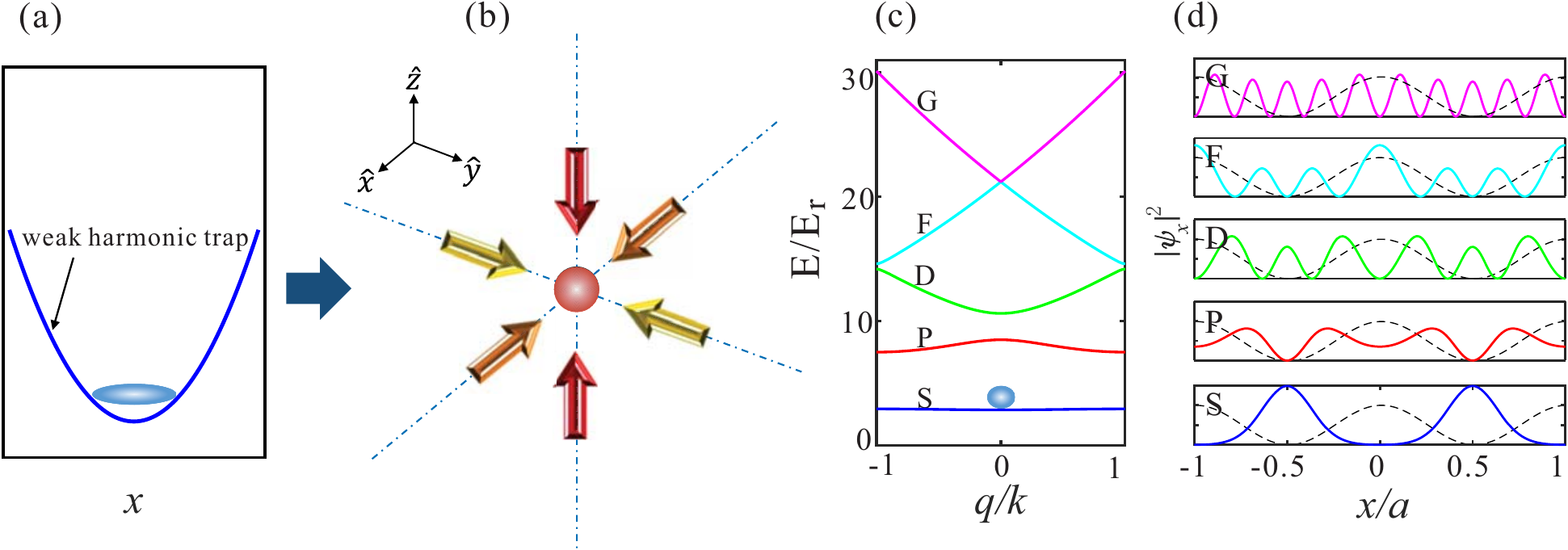}
  \end{center}
  \caption{The system before and after the preparation process. (a) Before the process, the atoms are confined in a weak
  harmonic trap. (b) After the preparation, the atoms are transferred into a desired band of a 3D OL. (c) The band structure of 1D OL versus quasi-momentum $q$ in the first Brillouin zone for $V_0=10E_r$. (d) From the bottom to top, the solid lines are spatial density distributions of different Bloch states (S,P,D,F,G) with $q=0$. The dashed line is the spatial distribution of the lattice potential.}\label{fig:f1}
\end{figure}
\section{The shortcut loading method}\label{Method}
\subsection{The idea of shortcut}\label{shortcut}

We consider the general situation for transferring atoms into an OL. Before the preparation, atoms are confined in a weak
harmonic trap $V_{harm}=\frac{1}{2} m (\omega_{x}^{2}x^{2}+\omega_{y}^{2}y^{2}+\omega_{z}^{2} z^{2})$ with the initial wavefunction $\left| {\psi_i} \right\rangle=|p=0\rangle$, as shown in Fig.~\ref{fig:f1}(a), where $m$ is the atom mass, $\omega$ is the trap frequency and $p$ the atomic momentum. The loading process then transfers atoms into a target state of an OL. The lattice is constructed by a set of laser beams with electric field amplitude $\vec{E_i}$, whose potential can be written as
\begin{equation} \label{e1}
V(\vec{r})\propto-\sum_{i,j}\vec{E}_i\cdot\vec{E}_j\cos((\vec{k}_i-\vec{k}_j)\cdot\vec{r}+(\alpha_{i}-\alpha_{j})),
\end{equation}
where $k_{i}=\lambda_{i}/(2\pi)$ is the wave number, $\lambda_i$ is the wavelength and $\alpha_{i}$ is the initial phase of laser beam $i$. For a cubic lattice, we can assume $k_j=-k_i$ and $i=x,y,z$, as shown in Fig.~\ref{fig:f1}(b). When neglecting atom-atom interactions because loading time is very short and for the lattice laser sufficiently far detuned, the single-atom Hamiltonian in OL is given by $(\hbar =1)$, %
\begin{equation}
\hat{H}=\frac{\hat{p}^{2}}{2m}+V(\vec{r}).
\label{e1a}
\end{equation}
According to the Bloch's theorem, the eigenstates of the Hamitonian $\hat{H}$ can be expressed as
$\left|{n,\vec{q}}\right\rangle=u_{n,\vec{q}}(\vec{r})e^{i\vec{q}\cdot \vec{r}}$, with the index of the energy band $n=1,2,3...$  and the quasi-momentum $\vec q$.

We first consider the 1D case for simplicity. The potential can be expressed as  $V(x)=\frac{V_0}{2}(1+\cos{2kx})$, where $V_0$ is the lattice depth (here the harmonic trap is ignored during the preparation process because it is small compared with the OL potential). The Bloch states can be written as
\begin{equation}
\left|{n,q}\right\rangle =\sum_{\ell}c_{n,\ell}\left|{2\ell k+q}\right\rangle.
\label{e2}
\end{equation}
This target state $\left| {\psi_a} \right\rangle$ can be decomposed over a reduced basis of plane waves $| 2{\ell}k+q\rangle$.
In the quasi-momentum space, for a pure Bloch state at $q=0$, the parity is given by $\Omega=\sum_{\ell}\left|c_{n,\ell}-c_{n,-\ell}\right|^2/4$, where $\Omega=1$ stands for a state with odd parity and $\Omega=0$ even. As shown in Fig.~\ref{fig:f1}(c), the Bloch state with $n=1,3,5...$ correspond to the S-, D-, G-bands with even parity,  and $n=2,4,...$ to the P- and F-bands with odd parity ( $V_0=10E_r$ where $E_r$ is one-photon recoil energy $E_r=k^2/(2m)$). The corresponding wave functions for the different Bloch states (S,P,D,F,G) with $q=0$ are also shown in Fig.~\ref{fig:f1}(d).

To achieve fast loading we will apply a $m$-step preloading sequence on the initial state $\left|\psi_i\right\rangle$ before switching on the lattice with the optical depth $V_0$.   The state after the preloading sequence $\left| {\psi_f} \right\rangle$ is given by:
\begin{equation}
\centering|\psi_f\rangle= \prod_{j=m}^{1} \hat{U}_j|\psi_i\rangle,
\label{sequence}
\end{equation}
where $\hat{U}_j=e^{-i\hat{H}_j t_j}$ is the evolution operator of the $j^{th}$ process. For the target state $\left| {\psi_a} \right\rangle$,  the parameters $\hat{H}_j$ and $t_j$ can be determined via maximizing the fidelity
\begin{equation}
\zeta=|\langle \psi_a|\psi_f\rangle|^2.
\label{fidelity}
\end{equation}
When $\zeta=1$ all the atoms would be prepared in the state $\left| {\psi_a} \right\rangle$. $1-\zeta$ describes the difference between the achieved atomic state $|\psi_f\rangle$
 and the target state $\left| {\psi_a} \right\rangle$. In the other word,  the deviation rate $N_e$ is:
\begin{equation}
\centering N_e=1-|\langle \psi_a|\psi_f\rangle|^2 .
\end{equation}
Our goal is to properly choose $\hat{H}_j$ and $t_j$ so that $N_e$ is small enough
to be neglected in the experiment.

\subsection{Calculating the time sequences}

 \begin{figure}[t]
  \begin{center}
	\includegraphics[width=0.9\textwidth]{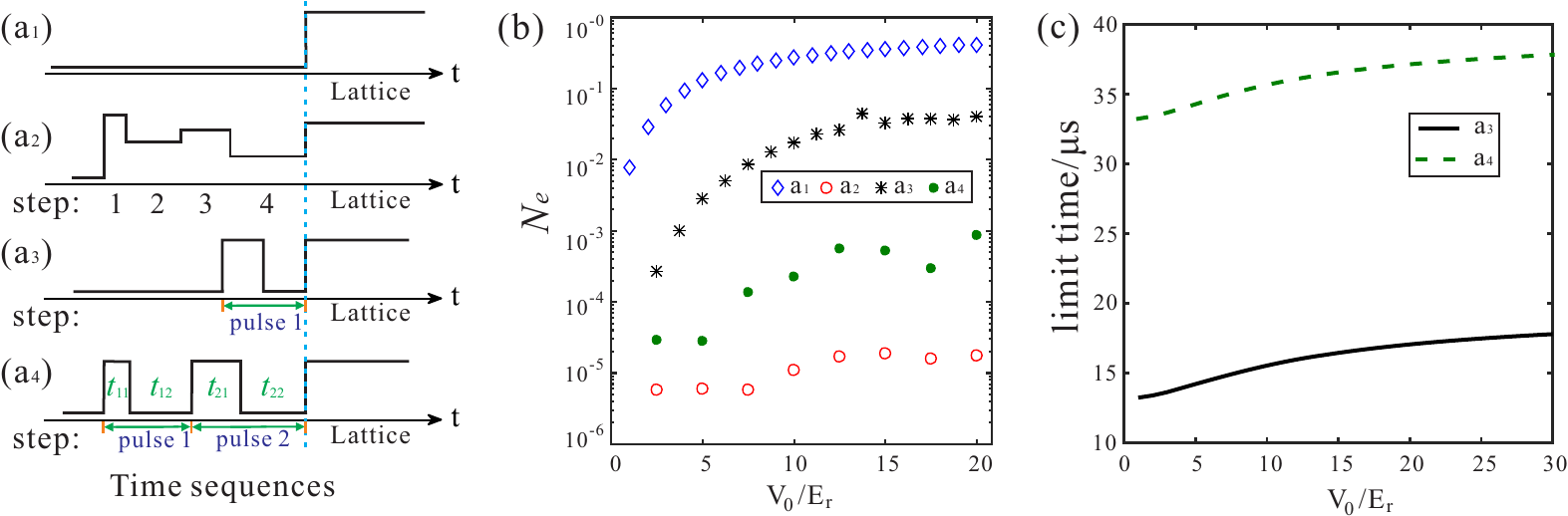}
  \end{center}
  \caption{(a) Four different time schemes for non-adiabatic loading into the ground (S) band. The OL turns on abruptly (a1); a four-step preloading sequence with both the potential depth and the duration of each step set as free parameters (a2); a one pulse (a3) and a two pulses (a4) preloading sequence with fixed potential depth and variable duration in each step. (b) Deviation rate $N_e$ for different lattice depths, where the blue diamonds, red circles, black stars and green points corresponding to (a1)-(a4), respecively. A logarithmic scale for the vertical axis is used. (c) Typical loading time for one pulse (a3)(solid line) and two pulses(a4) (dash line).}\label{fig:f2}
\end{figure}

One obvious choice for each $\hat{H}_j$ is to take the Hamiltonian corresponding to the interaction of atoms with a standing wave with the same periodicity as the OL. For this purpose, the power of the same laser as the one used for the final lattice loading is simply adjusted and each Hamiltonian $\hat{H}_j$
is obtained after substitution of $V_0$ by the new lattice depth $V_j$. More precisely, as $\hat{H}_j$ has spatial
periodicity, we get its eigenstates by solving the equation
$\hat{H}_j|n,q,V_j\rangle=E_{n,q}|n,q,V_j\rangle$~\cite{JPB02}, where $E_{n,q}$ is the corresponding eigenenergy. We use the notation
$|n,q,V_j\rangle$ for denoting the Bloch states for a $V_j$ lattice depth. Since only states with $q=0$ are initially populated, no other quasimomenta can be populated during the sequence of pulses. Then the state of the system can be written in the momentum eigenstates basis $| 2{\ell} k+q\rangle$, independent on the
potential depth $V_j$ and the
evolution operator can be written as the following matrix:
\begin{equation}
\hat{U}_j(V_j,t_j)=\hat{C}(V_j)\hat{E}(V_j,t_j)\hat{C}(V_j)^{\dagger},
\end{equation}
where $\hat{C}(V_j)$ is the unitary matrix of transition between the Bloch states basis and the momentum eigenstates basis with matrix elements
\begin{equation}
\hat{C}(V_j)_{{\ell}n}=\langle 2{\ell} k+q|n,
q,V_j\rangle,
\end{equation}
 and $\hat{E}(V_j,t_j)$ is a diagonal matrix with elements
 \begin{equation}
\hat{E}(V_j,t_j)_{nn}=\exp(-iE_{n,q}(V_j)t_j).
\end{equation}
Because of the simple form of the potential, it is easy to obtain these matrices, from which the wave function's evolution can be calculated optimally for a specific target state.
We can obtain the values of $V_j$ and $t_j$ by optimising for the specific target state using for example a  gradient descent algorithm.

Let's start with four steps, for which depth $V_j$ and duration time $t_j$ ($j=1,2,3,4$) are independently adjusted as shown in Fig.~\ref{fig:f2}(a2). Optimising for maximum fidelity (minimal deviation rate $N_e$ and constraining $t_j$ between $0\mu s$ to $50\mu s$ and $V_j$ from $0 E_r$ to $30 E_r$) we find $N_e$ to be in the rage of $10^{-5}$ or smaller for lattice depth of up to $30 E_r$ (Fig.~\ref{fig:f2}(b)).

Next we turn to a simpler control: keeping the lattice strength $V_j=V_0$ for $j=1,3$ (fixed to the final lattice potential $V_0$), and $V_j=0$ for $j=2,4$, the times $t_j$ being free parameters. This makes the sequence very easy to implement experimentally. A series of on and off pulses can be combined to a pulse sequence of length $m$, where the $j^{th}$ component is composed of a duration $t_{j1}$ where the OL is on and an interval $t_{j2}$ where the OL is off (Fig.~\ref{fig:f2}(a4)).  To obtain an optimised shortcut scheme we have to find the proper time sequences so that the fidelity $\zeta = {\left| {\left\langle {{\psi_f}} \right.\left| {{\psi_a}} \right\rangle } \right|^2} \to 1$. From the green points in Fig.~\ref{fig:f2}(b), we can see the deviation rate is still lower than $0.1\%$ for all lattice depths. If we only use one optimized pulse, as shown in Fig.~\ref{fig:f2}(a3), the fidelity is lower than $99\%$ for most of the considered OL depths, but still much better then just switching on the OL (Fig.~\ref{fig:f2}(a1)).  On the other side using more pulses, extending the sequence Fig.~\ref{fig:f2}(a4) there is still a small improvement. In addition, the typical times of the loading process under a two-level model approximation are given in Fig.~\ref{fig:f2}(c) for one pulse and two pulses. The improvement of fidelity comes at the expense of the loading time.

In the rest of our study we choose the simple scheme illustrated in Fig.~\ref{fig:f2}(a4).

\subsection{Experimental methods to probe the final state}

All our measurements are done in absorption imaging after 31 ms of \textit{Time of Flight} (TOF). The image thereby reflects the momentum of the atoms after the release from the optical lattice.

If we switch off the lattice abruptly (non-adiabatic switch off (NAS)),  we project the wave function  of the atoms in the lattice onto its momentum states.  If there is coherence in the trapped wave function, one observes diffraction peaks after time of flight.

If we switch off the lattice adiabatically, then we map the atom wave function in different bands to different momentum components.  This so called \textit{Band Mapping} (BM) ~\cite{Muller, EsslingerMap1,SpreeuwMap,EsslingerMap2} allows to investigate in which bands the atoms reside.  In addition the distribution of the atoms inside the mapped Brillouin zone allow to measure the distribution of quasi momenta. In our experiments the 'adiabatic switch off' is accomplished by exponentially ramping down the OL lattice potential in the form $e^{-t/\eta}$ where a characteristic decay time $\eta=100\mu s$ for a total length of 500 $\mu$s.

\subsection{Experimental measurement for loading atoms into $S$ band}
\begin{figure}[t]
  \begin{center}
	\includegraphics[width=0.8\textwidth]{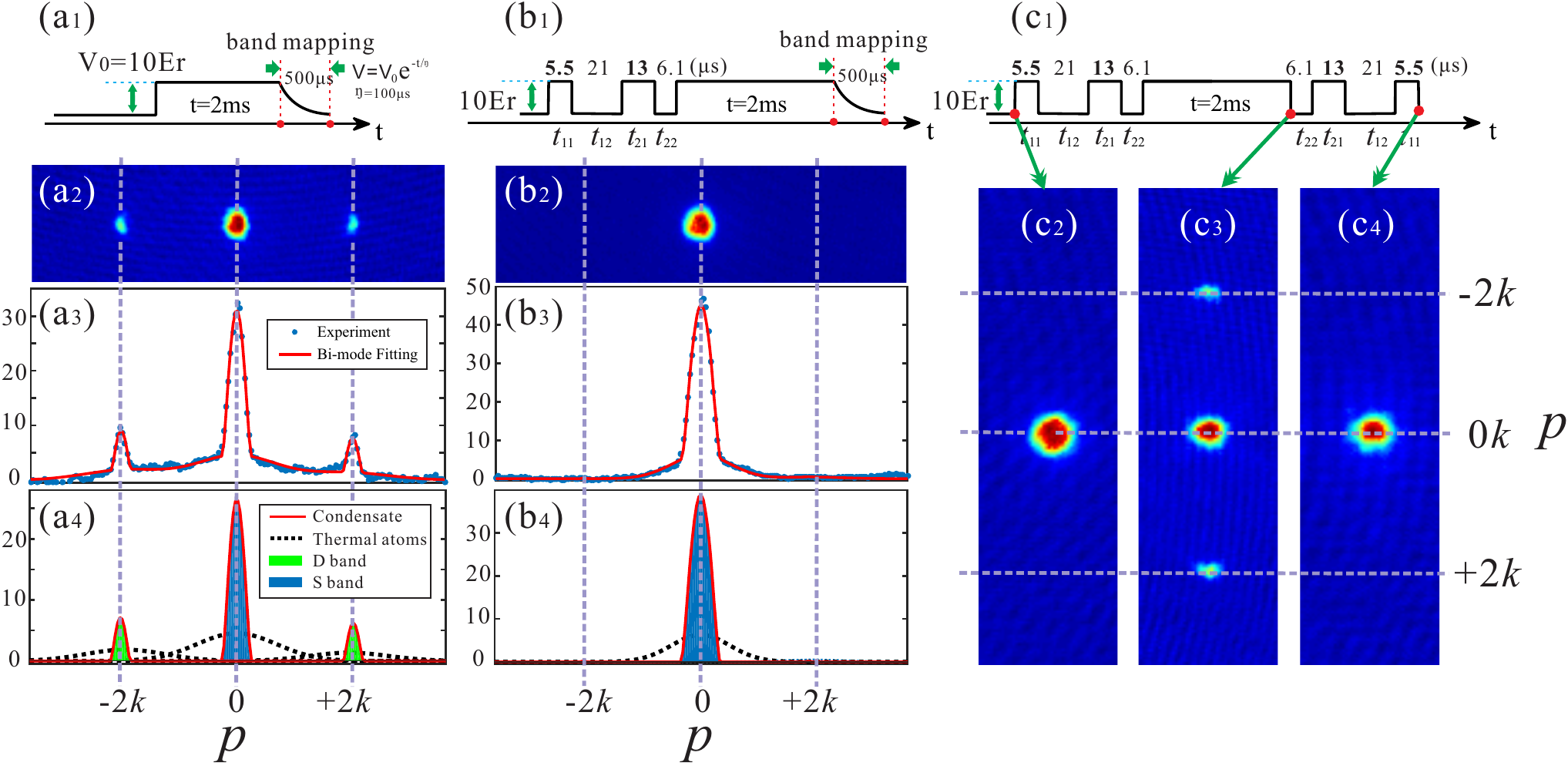}
  \end{center}
  \caption{The experimental demonstration of loading atoms into the S-band. The measured loading fidelity without (a) and with (b) shortcut method. The used time sequence, and the absorption image after band mapping (BM) are given in (a1) and (a2), respectively. We integrate the image along the $z$ direction and fit the atom distribution (blue points) by a Bi-modal function(red line) (a3,b3). The atom numbers for $S$(blue area) and $D$(green area) band can be gotten in (a4,b4). (c) After the shortcut and holding in the OL, we use a reverted pulse sequence to transfer the atoms back to the original state (c1). (c2-c4) show absorption images with non-adiabatic switching off (NAS) at the initial time, before and after using the reverted pulse sequence, respectively.}\label{fig:f3}
\end{figure}

To demonstrate our shortcut approach, we prepare a nearly pure BEC of about $1.5\times10^5$ $^{87}$Rb atoms in a hybrid trap which is formed by overlapping a single-beam optical dipole trap with wavelength 1064nm and a quadrupole magnetic trap. The resulting potential has harmonic trapping frequencies $(\omega_x,\omega_y,\omega_z)=2\pi\times(28,55,65)$Hz, respectively, and a temperature of about $60nK$. The lattice is implemented by a standing wave created by two counter-propagating laser beams along the $x$-direction, with the lattice constant being $\lambda/2=426$nm, the recoil energy $E_r$ being $3.16k\text{Hz}$.

We start with sequence Fig.~\ref{fig:f3}(a1), switching the OL with $V_{0}=10E_{r}$ on abruptly and hold the atoms in the lattice for $t=2\ ms$.  The absorption image after BM (Fig.~\ref{fig:f3}(a2)) shows a significant fraction of atoms at momentum $\pm 2 k$, which means that they are in excited bands (here the D-band).  We integrated this image along the direction perpendicular to the $\hat{x}$-axis and fit the experimental data points by three Bi-modal functions, as shown in Fig.~\ref{fig:f3}(a3) and (a4), respectively. The bi-modal function contains a Gaussian form that represents thermal atoms and an inverted parabolic function that denotes the condensate in S- and D-band. The blue and green area size equal to the atom numbers for S- and D-band, respectively. The measured fidelity from Fig.~\ref{fig:f3}(a3) and (a4) is $\zeta=72.6\%$.

We then realized an optimised 2 pulse shortcut sequence $(t_{11}, t_{12}, t_{21}, t_{22}) = (\textbf{5.5}, 21.0, \textbf{13.0}, 6.1) \mu s$ with the fixed depth $V_{0}=10E_{r}$ (Fig.~\ref{fig:f3}(b)). Employing band mapping we verify that the atoms are distributed in the first Brillouin zone that means atoms occupy $\left|S\right>$ band. Similar to above, we can measure the fidelity of our final state prepared by the shortcut method, and obtain $\zeta=99.2\%$.

To further show that the coherence is not destroyed in the transfer process, we first hold the atoms in the OL for 2 ms and then use two additional inverted pulses to transfer the atoms back to the original state $|\psi_i\rangle$, as shown in Fig.~\ref{fig:f3}(c). We study the state of the atoms by a non-adiabatic switching off (NAS). The images obtained with the initial condensate, the atoms loaded in the OL, and after two additional inverted pulses are shown in Fig.~\ref{fig:f3}(c2) to (c4), respectively. In Fig.~\ref{fig:f3}(c3), we can see the interference peaks, similar to the familiar pattern observed in adiabatic loading experiments, which indicates a successful loading without significant excitation and heating. Comparing Fig.~\ref{fig:f3}(c2) and (c4), we know there is little heating or disturbing effect on our BEC, which proves the effectiveness of our 'preparing' process of the ground state of OL~\cite{Liu}.

\subsection{Robustness analysis}
\begin{figure}[b]
  \begin{center}
	\includegraphics[width=0.95\textwidth]{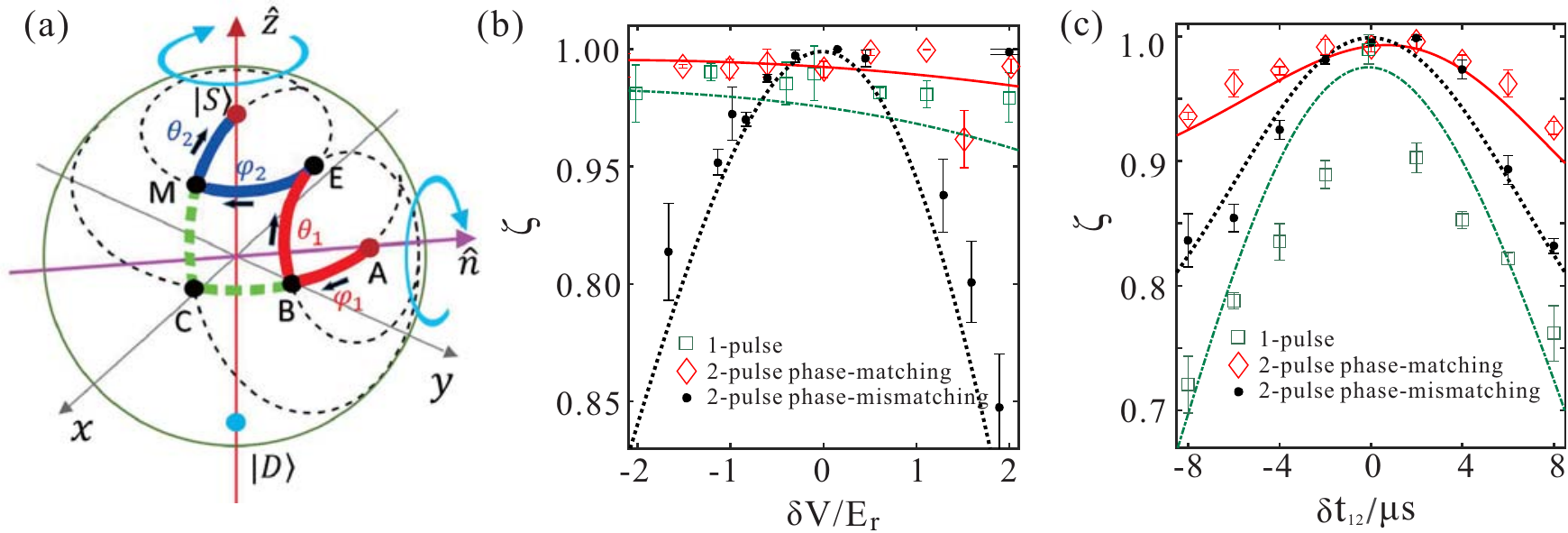}
  \end{center}
  \caption{(a): The track of the shortcut loading process on the Bloch sphere. The axis $\hat z$ and $\hat n$ are the rotation axis for the states corresponding to the OL being on and off, respectively.  The dashed lines are for one pulse sequence, demonstrated from point A to the medium state point C, then the final state $| S \rangle $. The case of two pulses is shown with red and blue solid lines, which start from the initial state point A, go through the middle states points B, E, and M, and get to the final state $| S \rangle $. The variation of the fidelities with $\delta V$ (b) and $\delta t_{12}$ (c) for the different pulse sequences ($V_{0}=10E_r$). Theory curves are shown as the dash-dotted (one pulse), dotted (two pulse sequences process without phase-matching) and solid (two pulse sequences process with phase-matching) curves, while the corresponding experimental datas are shown as square, dotted and diamond points, respectively.
  }\label{fig:f4}
\end{figure}

In order to analyse the robustness of the pulse sequences, we use a two-level model approximation to draw the trajectory of the evolution on the Bloch sphere. Considering the parity of the bands, a two-level model ($\left| S \right\rangle$, and $\left| D \right\rangle$ with corresponding eigenvalues $E_S$ and $E_D$) is suffitient when the OL depth is low. As shown in Fig.~\ref{fig:f4}(a), we choose the aimed state as the S-band with zero quasi-momentum. The polar axis (${\hat z}$ axis) represents the Bloch state $\left| S \right\rangle$ ($\left| D \right\rangle$) in the positive (negative) direction. Consequently, the initial plane wave represented by axis $\hat n$ can be set as $\left| {\psi_i} \right\rangle = {\left( {\cos \frac{\beta }{2},\sin \frac{\beta }{2}} \right)^T}$, where $\beta$ is the angle between axis ${\hat z}$ and ${\hat n}$. Obviously, when the pulse is imposed, the action  can be seen as an counterclockwise rotation of $\varphi_s$ around axis ${\hat z}$ (the state $\left| S \right\rangle$) for a vector in the Bloch sphere. Likewise, during the time interval with the pulse being off, it is equivalent to an counterclockwise rotation of $\theta_s$ around the axis $\hat n$ (the plane wave with zero momentum) for a vector in the Bloch sphere.

As we can see in Fig~\ref{fig:f4}(a), the path (from point A to B, E to M) for ${\hat U}_{j1}$ caused the change in the phase between Bloch bands. On the other hand, the path (from point B to E, M to S) for ${\hat U}_{j2}$ mainly caused the change in the proportion of Bloch bands. We can obtain many different sequences for the same $\left| {\psi_a} \right\rangle$ with $\zeta\to1$. From the analysis of the trajectories on the Bloch sphere, we find that if the track is symmetric, it is the most robust. For the track in Fig.~\ref{fig:f4}(a), if
\begin{eqnarray}
{\varphi_1} = {\theta_2}\ \text{and}\ {\theta_1} = {\varphi_2} \label{phase}
\end{eqnarray}
and ${\rm d}^2\zeta/{\rm d}\gamma^2$ is the minimum, where $\gamma$ are the parameters in experiment such as $t_{ij}$ and $V_0$, which reminds us at a phase matching condition.

In Fig.~\ref{fig:f4}(a), it is clear that this matching conditions have mirror symmetry about the center of the whole loading path. Therefore, we should choose proper rotation angles to make the fidelity maximum and  get the highest robustness. Considering the influence of higher bands, the sequence with phase matching in the two-level approximation will have to be corrected in order to satisfy the multi-level condition.

The variation of the fidelity $\zeta$ with respect to the pulse amplitude (mismatch with the lattice depth) is shown in Fig.~\ref{fig:f4}(b).  The experimental results are in good agreement with the theoretical predictions. The diamond points, which represents the shortcut time sequence considering the phase matching and multi-level correction, is the most robust. There is less than $0.2\%$ variation for $\delta V=0.5E_r$. On the contrary, there is $2\%$ variation when we do not consider the phase matching as shown in the dotted points. Furthermore, although both one pulse and two pulses sequences satisfy the phase matching conditions, Fig.~\ref{fig:f4}(c) indicates that the robustness and fidelity of one pulse sequence is a lower than two pulse sequence with respect to variation of the time.

\section{Loading atoms into higher bands}\label{Higher}

The above shortcut method can be adapted to load atoms into excited bands in an OL. Since the initial state is of even parity and the parity of wave function remains unchanged during the pulse sequence, we can easily load atoms into higher bands of even parity such as D- and G-band. By adding a  shift of the lattice phase we can change the parity to transfer atoms into odd parity bands such as the P- or F-bands.

\subsection{Loading atoms into D-band}\label{D band}
\begin{figure}[t]
  \begin{center}
	\includegraphics[width=0.8\textwidth]{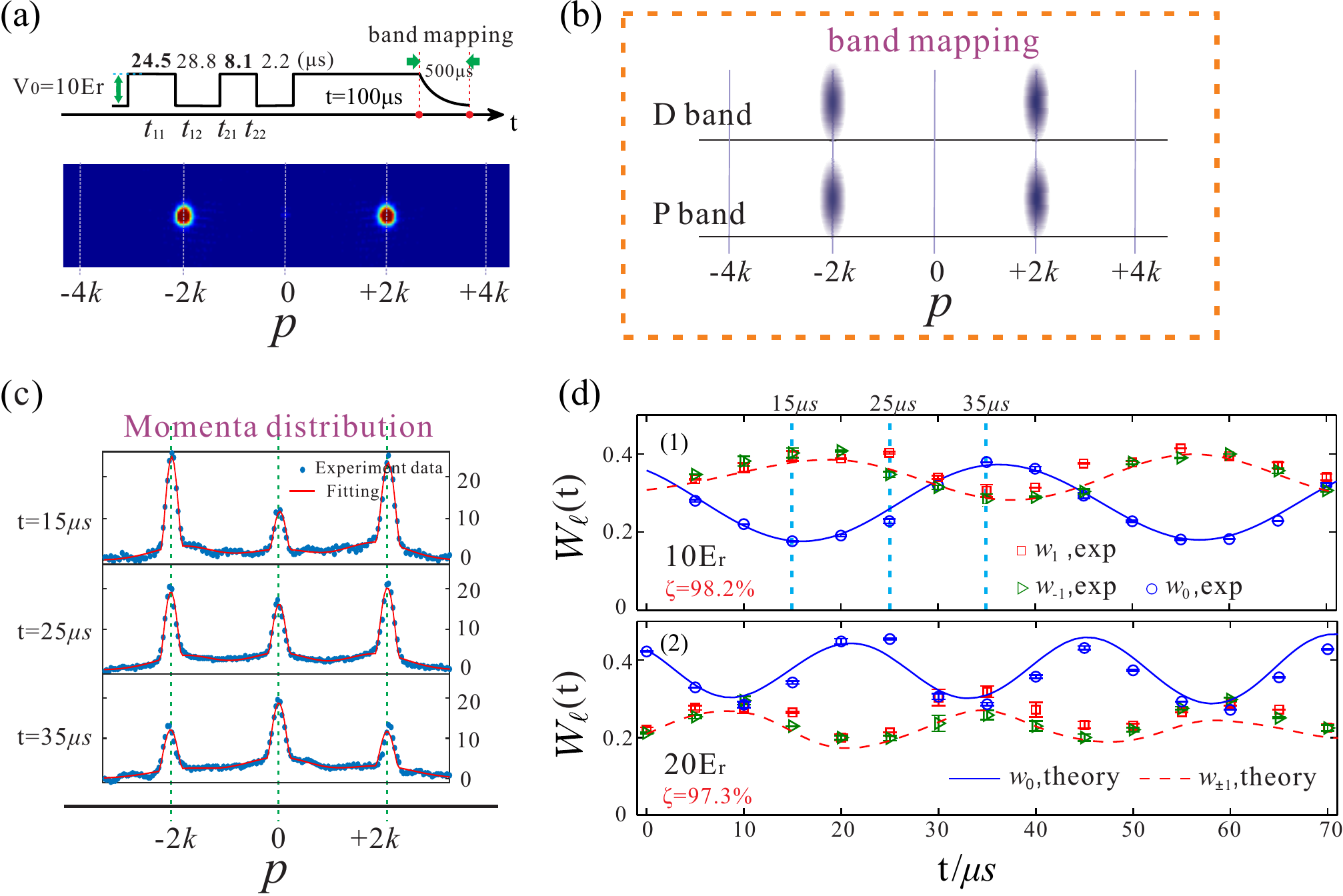}
  \end{center}
  \caption{ The demonstration of loading atoms into the D-band. (a) Time sequence and the experimental image after BM. (b) The BM diagram for atoms in the P- and D-band with zero quasi-momentum. (c) Integration of the momentum distributions in the vertical direction of the OL for experimental images after NAS. The blue points are the experimental data, and the red lines are results from the bi-modal fit. From the fit we can extract the numbers of atoms in $0$ and $\pm2k$. The three images correspond to holding time $15\mu s$, $25\mu s$ and $35\mu s$, respectively. (d) The oscillations of the relative population $W_\ell(t)=N_\ell(t)/N$ $(\ell=0, \pm1)$ with the holding time $t$ for case (1) $V_0=10E_r$ and (2) $V_0=20E_r$. We show the measured values of $N_0(t)/N$ (blue circles), $N_1(t)/N$ (red squares), and $N_{-1}(t)/N$ (green triangles) from the experiments, and the corresponding theoretical calculations as the blue solid lines and red dashed lines, respectively.
  }\label{fig:f5}
\end{figure}

Similar to the loading method into the S-band, we can numerically maximize the fidelity $\zeta$
to obtain the time sequence to load atoms into the D-band. The time sequence is $(t_{11},t_{12},t_{21},t_{22})=(\textbf{24.5},28.8,\textbf{8.1},2.2)\mu s$ for $V_{0}=10E_r$, and the experimental image by BM is shown in Fig.~\ref{fig:f5}(a). However, we could not get the fidelity from this image because both D- and P-band atoms are distributed at $\pm 2 k$, as illustrated in Fig.~\ref{fig:f5}(b).

We can measure the loading fidelity by the oscillations of the relative population in different momenta from the images by NAS. After the preparation process, atoms are in state
\begin{eqnarray}
|\psi _{f}\rangle={\prod\limits_{j = m}^1 {{\hat U}_{j2} {\hat U}_{j1}} } \left| {p_x=0} \right\rangle \equiv \sum_{n}f_{n}|n,0\rangle, \label{psil}
\end{eqnarray}%
At holding time $t$ in the OL, the atomic state becomes $|\psi _{t } \rangle =\sum_{n}f_{n} e^{-iE_{n ,0}t }|n,0\rangle $, where $E_{n,0}$ and $|n,0\rangle$ are the eigen-energy and eigen-state of $H_{x}$, $H_{x}|n,0\rangle =E_{n,0}|n,0\rangle$. Therefore, the number of atoms $N_{\ell}(t)$ in the state $|p_{x}=2\ell k\rangle$ ($\ell=0, \pm1$) at time $t$ is given by $N_{\ell}(t )=N|\langle p_{x}=2\ell k|\psi _{t}\rangle |^{2}$, and satisfies
\begin{eqnarray}
W_\ell(t)\equiv\frac{N_{\ell}(t )}{N}=\left\vert \sum_{n}f_{n}c_{n,\ell}e^{-iE_{n ,0}t }\right\vert ^{2},  \label{g}
\end{eqnarray}%
where $N$ is the total atom number and $c_{n,\ell}$ is defined in Sec.~\ref{shortcut}. Eq.(\ref{g}) shows that the atom number $N_{\ell}(t )$ oscillates with time $t $.

Fig.~\ref{fig:f5}(c) shows the momentum distribution along $\hat{x}$ direction extracted from experimental images obtained by NAS. By a bi-modal fit  we  obtain  $N_{\ell}$ and $W_\ell$ for different momentum states  $|p_{x}$=$2\ell k\rangle$. The experimental values $W_\ell(t)$ oscillate with time as shown in Fig.~\ref{fig:f5}(d) and are well described by the theoretical model Eq.(\ref{g}). From a fit to the experimental data, blue solid line for $W_0(t)$) and red dash line for $W_{\pm1}(t)$ (we set $W_1(t)=W_{-1}(t)$ in the theory calculations) we extract the fidelity of the preparation process:  $98.2\%$ for $10E_r$ and $97.3\%$ for $20E_r$ using the loading time sequence $(\textbf{17.2},25,\textbf{12.5},1.1)\mu s$. After loading, the lifetime of atoms in D-band can be measured~\cite{Zhai}.

\subsection{Loading atoms into G-band}\label{G band}

\begin{figure}[b]
  \begin{center}
	\includegraphics[width=0.8\textwidth]{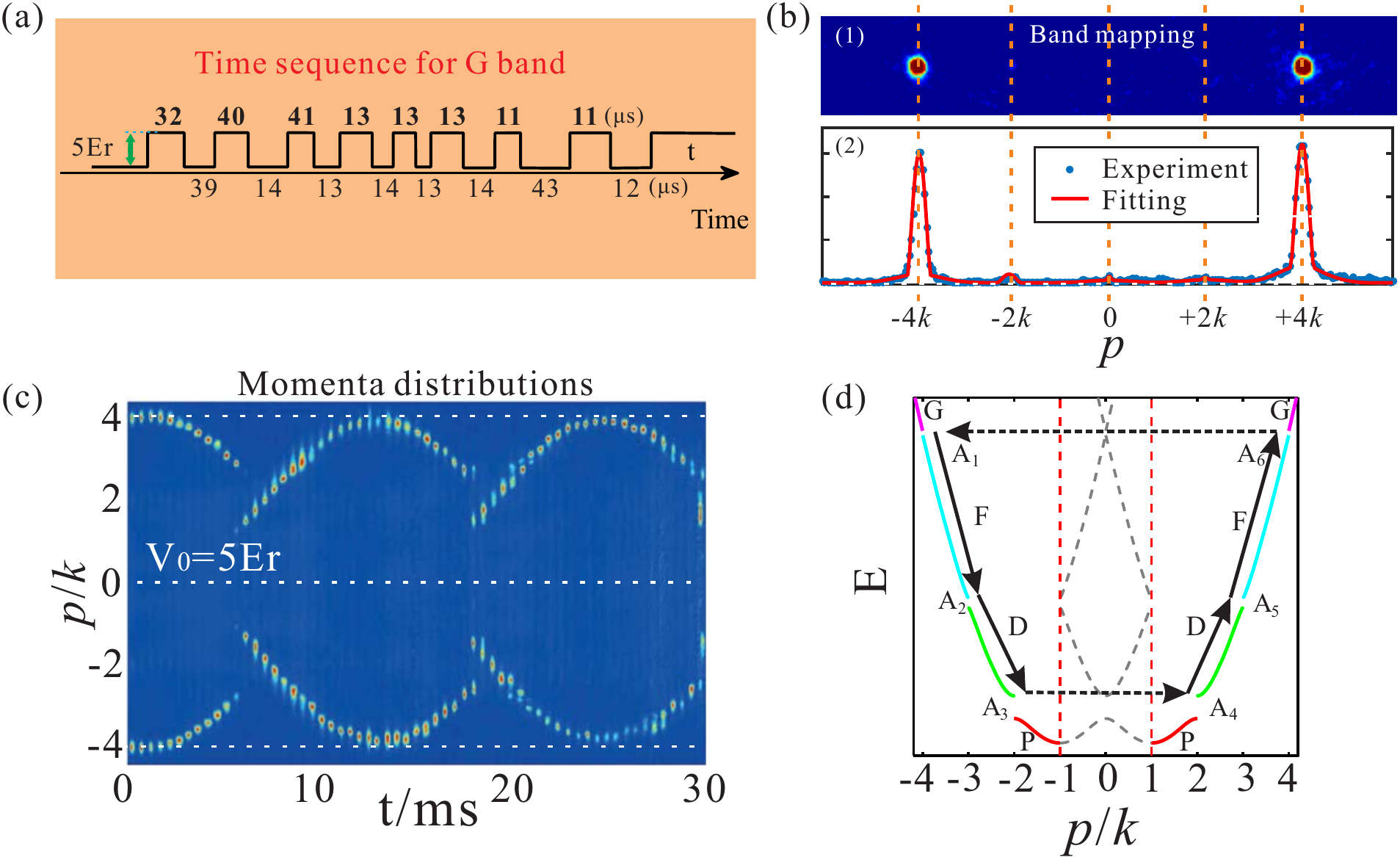}
  \end{center}
  \caption{ (a) The shortcut time sequence for loading atoms into G-band. The experimental image with BM  after holding 50$\mu s$ in an OL with $V_0=5E_r$ (b1) and its Bi-modal fit for integration along the vertical direction of OL (b2). (c) The dynamical oscillations of atoms after loading into G-band from the experiment by NAS method. (d) Schematic of extended Bloch bands (P, D, F and G).
  }\label{fig:f6}
\end{figure}

For even higher bands such as the G-band, theoretical calculations sugest a fidelity $\zeta=99.2\%$ for a $5E_r$ deep lattice using an eight pulses sequence as: $(\textbf{32},39,\textbf{40},14,\textbf{41},13,\textbf{13},14,\textbf{13},13,\textbf{13},14,\textbf{11},43,\textbf{11},12)\mu s$. The experimental image by BM, and its momentum distribution is given in Fig.~\ref{fig:f6}(b). We could not get the fidelity from the image by BM because both G- and F-band atoms are distributed at $\pm 4 k$.

Applying this shortcut method to load atoms into G-band at $q=0$, a dynamical oscillation is clearly visible in the images with NAS, as shown in Fig.~\ref{fig:f6}(c). This is best understood when looking at the corresponding extended band structure as drawn in Fig.~\ref{fig:f6}(d), where the energy gaps between different bands are marked with $A_s$ ($s=1,2,3,4,5,6$): After loading into G-band, the atoms fall down into the F-band due to the small gap between G and F bands~\cite{Wang}. During the G-band preparation process, we ignored the effect of the harmonic trap because the time of the shortcut is very short. However, when we observe the atoms in OL for a long time, the weak harmonic trap will affect the dynamics. Once the BEC is in the F-band, it continues to lose momentum while gaining potential energy from the harmonic confinement. This corresponds to the BEC traversing dynamically along the F-band from $A_1$, $A_6$ to $A_2$, $A_5$ in Fig.~\ref{fig:f6}(d). Once arriving at $A_2$ or $A_5$, the atoms face different dynamics depending on the lattice strength. If the lattice strength is small and the Bragg reflections at $A_2$ and $A_5$ are weak, the BEC will continue into the D-band by a Landau-Zener transition. After evolving along the entire D-band, the BEC comes to the band gap between D- and P-bands at $A_3$ and $A_4$. Due to the large gap between D and P bands all the atoms at $A_3$($A_4$) will be Bragg reflected to $A_4$($A_3$), without tunneling into the P-band. Afterwards, the BEC will reverse its dynamics by moving up in momentum from $A_4$, $A_3$ to $A_5$, $A_2$. It eventually arrives at $A_6$, $A_1$, completing half of an oscillating cycle. As illustrated in Fig.~\ref{fig:f6}(c), the oscillation period is about $24$ms.

The above atomic oscillation depends on the optical depth. When the OL is strong such as $V_0=15E_r$, the oscillation only exists within the F-band with a period of $17$ms. When the lattice strength is intermediate, for example $V_0=7.5 E_r$, there is a superposition of two kinds of oscillations: across both the F and D bands, and within the F-band~\cite{Hu}.

\subsection{Loading atoms into P-band}
\begin{figure}[t]
\begin{center}
\includegraphics[width=0.6\textwidth]{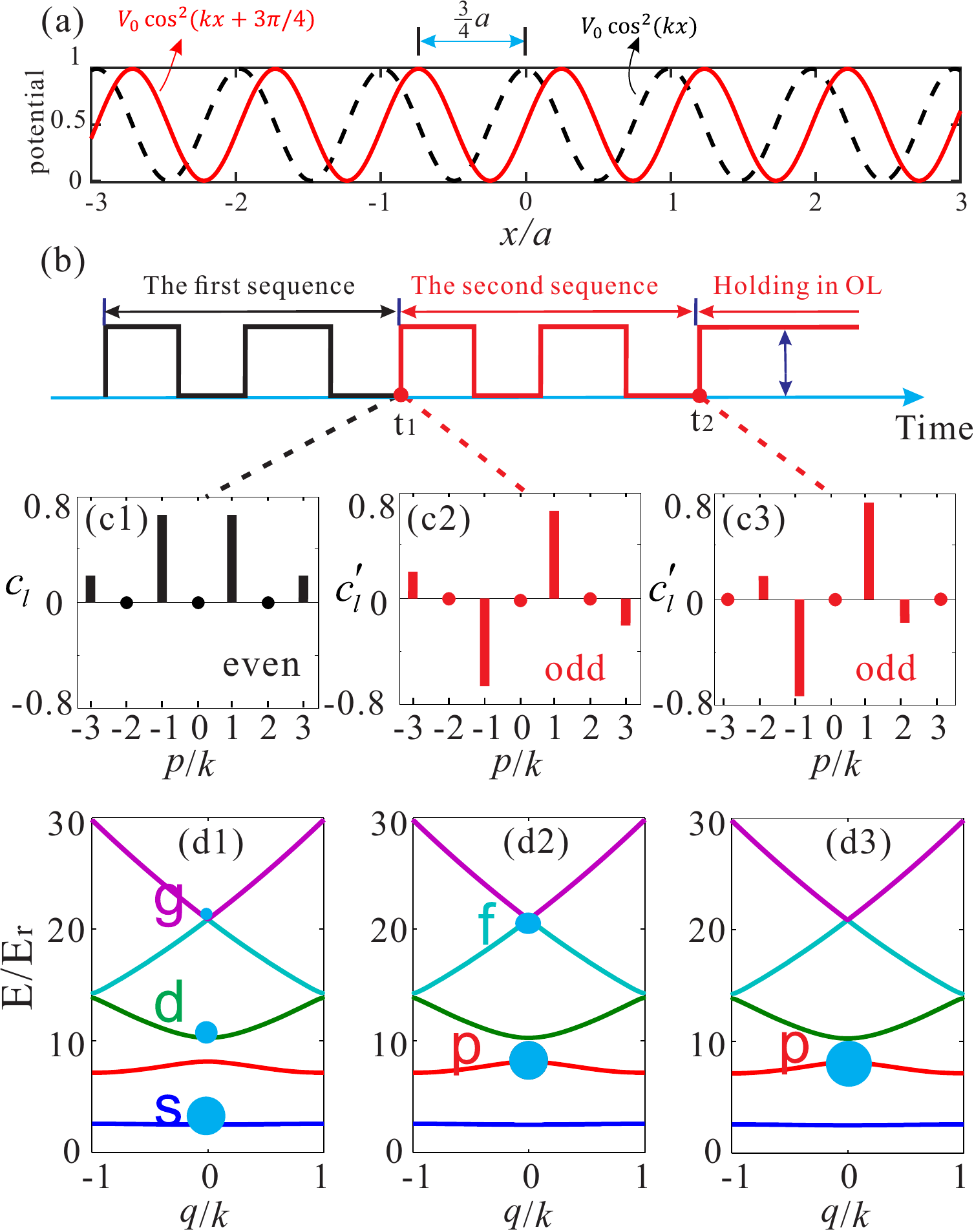}
\end{center}
\caption{Schematic to prepare atoms into the P-band.  (a) Two 1D lattices with a $3\pi/2$ phase shift. (b) Loading time sequences with the first series of pulses from $0$ to $t_{1}$ for $V_{even}(x)=V_0\cos
^{2}\left( kx\right)$ and the second from $t_{1}$ to $t_{2}$ for $V_{odd}(x)=V_0\cos^{2}\left(
kx+3\pi/4\right)$. (c1-c3) Superposition
coefficients $c_{\ell}$ right before $t_1$, $c^\prime_{\ell}$ at the
moment right after $t_1$ and at $t_2$, respectively. (d1-d3) The corresponding population distribution in the Bloch band.
}\label{fig:f8}
\end{figure}

\begin{figure}[t]
 \begin{center}
	\includegraphics[width=0.8\textwidth]{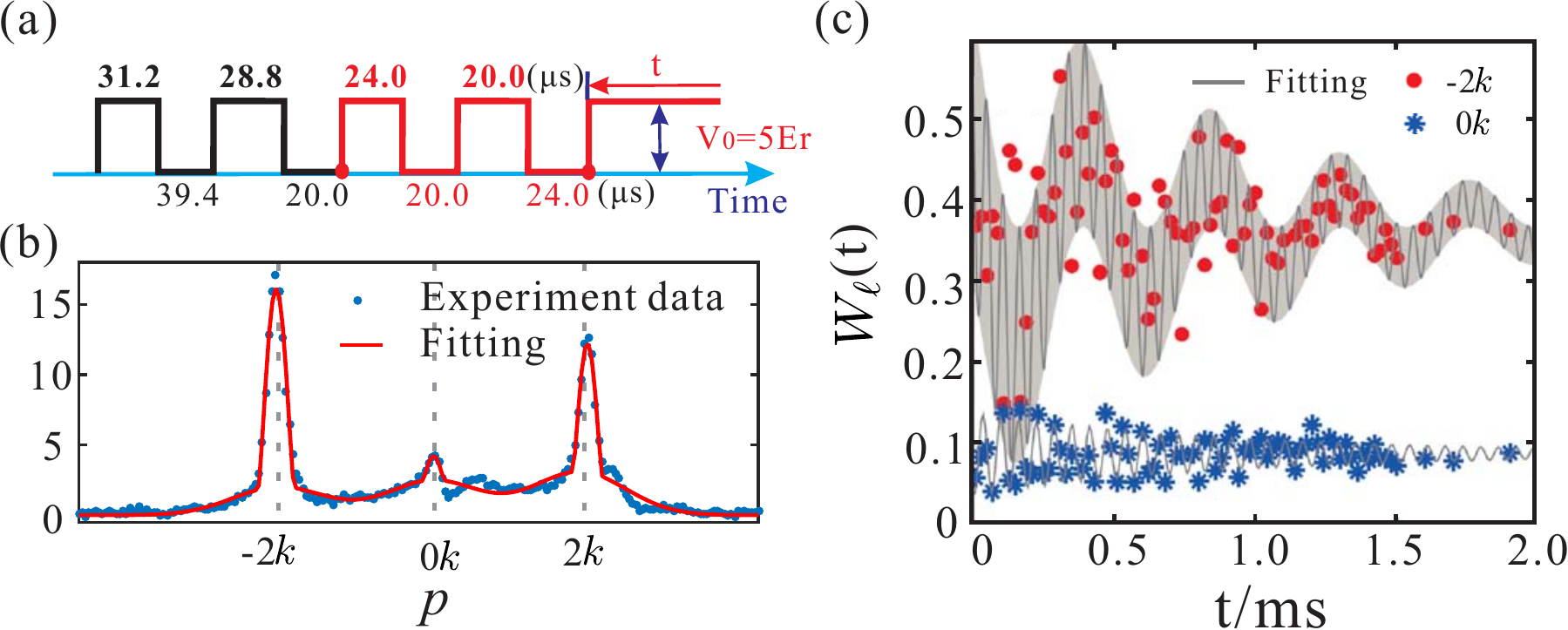}
  \end{center}
  \caption{ (a) Designed time sequence for loading atoms into P-band with $V_0=5E_r$. (b) The momentum distribution of absorption image by NAS along $\hat{x}$ direction. Experimental data (blue points) fitted by a bi-modal function (red line). (c) Population oscillations around momenta $-2k$(red points) and $0k$(blue stars). The solid lines are the fitting curves.
  }\label{fig:f9}
\end{figure}

The parity of quantum states in the P-band with $q=0$ is odd, $\psi(-x)=-\psi(x)$~\cite{Bloch2,speed,NP,shaken,GW}, one has to change the parity in order to load into the P-band. This can be done by a spatial shift of the OL. Our preparation process therefore consists of two series of pulses, as shown in Fig.~\ref{fig:f8}(a) and (b). In the first series of pulses from $0$ to $t_{1}$, the atom experiences a spatial potential  $V_{even}(x)=V_0\cos^{2}\left( kx\right)$. In the second series pulses from $t_{1}$ to $t_{2}$, the atom experiences a potential $V_{odd}(x)=V_0\cos^{2}\left(kx+3\pi/4\right)$. The coefficients $c_{\ell}$ ($c^\prime_{\ell}$) defined in Eq.(\ref{e2}) and the distribution in the Bloch bands at the same times are shown in Fig.~\ref{fig:f8}(c) and (d), respectively. At time $t_{1}$, all the components $c_\ell$ in the parity $\Omega$ would satisfy $c_\ell-c_{-\ell}=0$, as shown in Fig.~\ref{fig:f8}(c1) and the energy bands S, D and G... shown in Fig.~\ref{fig:f8}(d1).
However, from the view of the second series of pulses, by the lattice shift, the coefficient $c_\ell$ should be multiplied by a phase according to $l$, i.e. $c^\prime_\ell|\ell\rangle=c_\ell e^{i2\ell(3\pi/4)}|\ell\rangle$, and the relation between coefficients becomes $c^\prime_\ell-(-\ell)^{\ell}c^\prime_{-\ell}=0$. In our loading process, the first series of pulses ensure that coefficients $c^\prime_\ell$ with even $\ell$ are zero. At the beginning of the second series of pulses, the parity of states can be completely changed as shown in Fig.~\ref{fig:f8}(c2) and the corresponding energy bands P and F... are shown in Fig.~\ref{fig:f8}(d2). From $t_1$ to $t_2$, the parity is unchanged. So only $P$ band state is
populated at time $t_{2}$, as shown in Figs.~\ref{fig:f8}(c3) and (d3).

In the experiment, we use two acousto-optic modulators to form our designed pulse sequence with the frequency difference $\delta\omega=182.5$MHz which corresponds to a phase shift between two pulses series by $3\pi/4$. Four special pulses are used to transfer atoms into P-band for $V_0=5E_r$, where the first series is $(\textbf{31.2},39.4,\textbf{28.8},20.0)\mu s$ and the second is $(\textbf{24.0},20.0,\textbf{20.0},24.0)\mu s$, as shown in Fig.~\ref{fig:f9}(a). The momentum distributions of absorption image along $\hat{x}$ direction after NAS are shown in Fig.~\ref{fig:f9}(b). The momentum distribution nearly equals to zero at $0k$ and has significant peaks at $\pm 2k$. To obtain the loading fidelity of P-band, we can measure the oscillations of $W_\ell(t)$ $(\ell=0, -2)$ as shown in Fig.~\ref{fig:f9}(c), which is similar to D-band. By comparing the experimental data with the peripheral contour of the beating signal, we find the initial quantum state is $|\psi(t=0)\rangle=\sqrt{0.9}|P,q=0\rangle+\sqrt{0.05}|D,q=0\rangle+\sqrt{0.05}|S,q=0\rangle$~\cite{Hu}. The corresponding fidelity is about 90\% in P-band.

After loading atoms into P-band and holding for a longer time we can observe the quantum equilibration in dilute Bose gases~\cite{Niu1}. In a similar way, we can also transfer atoms into the F-band with two sets of standing wave pulses ${V_0}{\cos ^2}\left( {{k}x} \right)$ and ${V_0}{\cos ^2}\left( {{k}x{\rm{ + }}3\pi /4} \right)$~\cite{Wang}.

\section{Preparation and manipulation of superposition of Bloch states }\label{Manipulation}

Above, we presented a shortcut loading method to transfer atoms into one band at zero quasi-momentum. We now extend our scheme to load atoms into S-band with non-zero quasi-momentum and superpositions of band states. Furthermore, it can be used to construct $\pi/2$ pulse or $\pi$ pulse between S and D bands, and implement a Ramsey interferometer (RI) with motional states~\cite{Hu1}.

\subsection{Preparation of atoms in the S-band with non-zero quasi-momentum}

\begin{figure}[t]
 \begin{center}
	\includegraphics[width=0.9\textwidth]{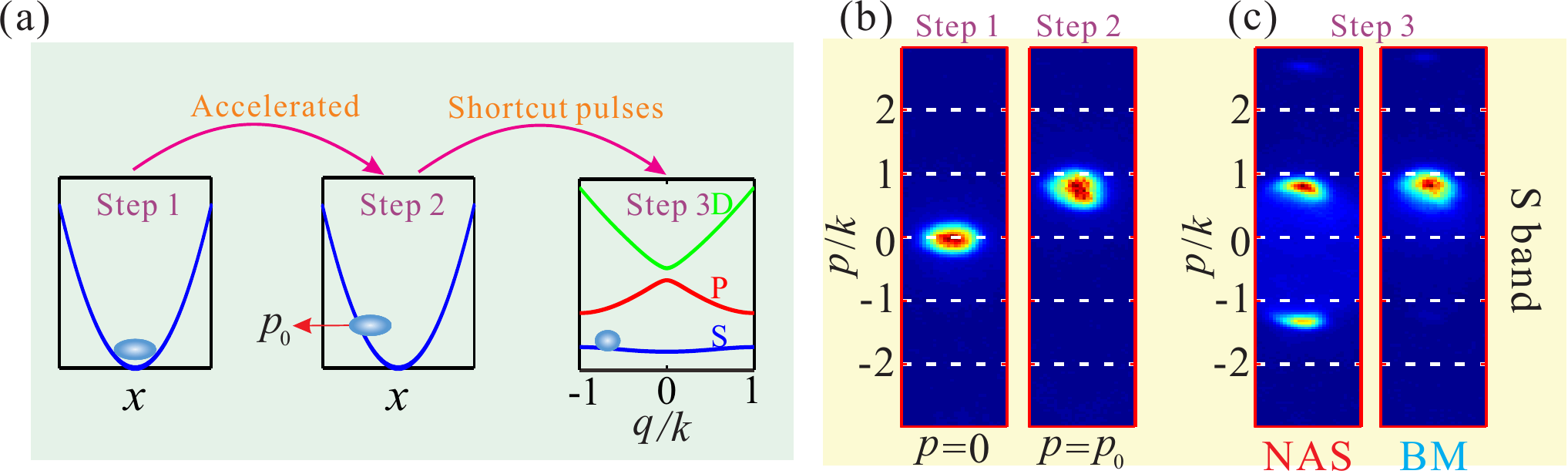}
  \end{center}
  \caption{Schematic diagrams for transferring atoms into the S-band with non-zero quasi-momentum (a): Step 1, the BEC in the ground state of a harmonic trap; Step 2, the atoms are accelerated to a momentum $p_0$; Step 3, using shortcut pulses, the atoms are transferred into the S-band of an OL with non-zero quasi-momentum. (b): The absorption images by NAS correspond to step 1 and step 2, respectively. (c) The images after step 3 by NAS (left) and BM (right) methods, respectively.}\label{fig:f10}
\end{figure}

If we want to load atoms into the S-band with non-zero quasi-momentum $q_0$, the BEC, initially with $p=0$ (Fig.~\ref{fig:f10}(a)), should be accelerated to obtain a momentum $p_0$. We can use a  magnetic field gradient provided by coils, to accelerate a BEC to momentum $p=-0.8 k$ within $2$ms. Immediately afterwards the designed pulse sequence is used to transfer atoms into the S-band of the OL at quasi-momentum $q_0$. The corresponding experimental absorption images by NAS after step 1 and step 2 are shown in Fig.~\ref{fig:f10}(b). After step 3, the final state $\left|S,q=-0.8k\right\rangle$ is shown in Fig.~\ref{fig:f10}(c). The momentum distribution as measured by NAS (left image) has significant peaks at $0.8k$ and $-1.2k$. The right image as obtained by BM shows a significant peak at $0.8k$, which verifies the effectiveness of the loading.

\subsection{Preparation of atoms in superposition of Bloch states}

\begin{figure}[b]
 \begin{center}
	\includegraphics[width=0.9\textwidth]{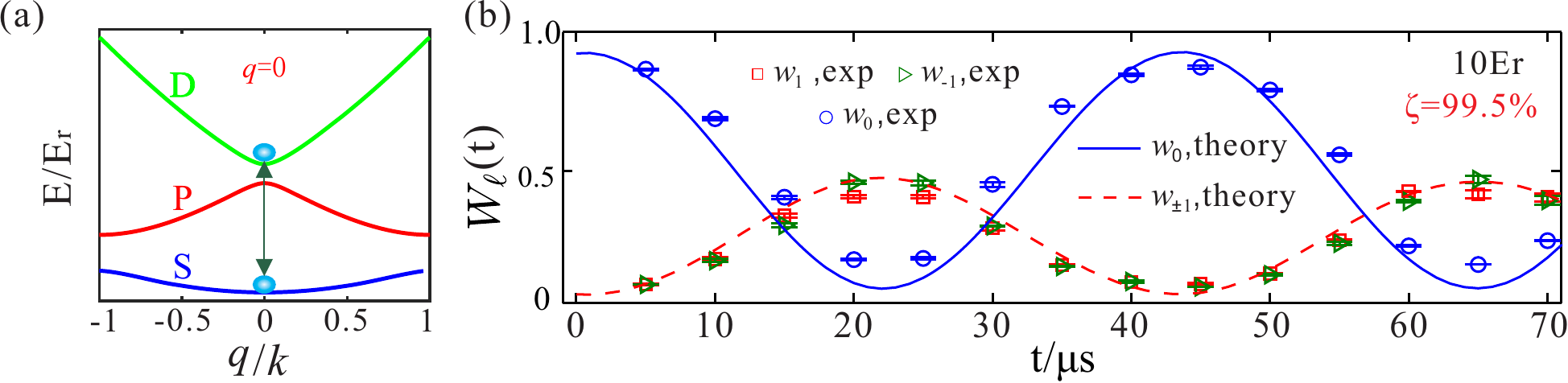}
  \end{center}
  \caption{Loading atoms into superposition states. (a) Schematic illustration of the superposition states $(|S\rangle +|D\rangle)/\sqrt{2}$ in OL with $q=0$ for $V_0=10E_r$. (b) The relative population as a function of holding time $t$ (similar to Fig.~\ref{fig:f5}(b)).}\label{fig:f11}
\end{figure}

We can also choose the target state as a superposition states, such as $|\psi _{a}\rangle =(|S\rangle +|D\rangle)/\sqrt{2}$, as illustrated in Fig.~\ref{fig:f11}~\cite{Zhai}. Using a shortcut time sequence as $(\textbf{30},6.4,\textbf{8.7},4.5)\mu s$ we extract a measured fidelity of $\zeta=0.995$ from the fits to the measured $W_\ell(t)$ ($\ell=0,\pm 1$) as a function of the holding time $t$.

\subsection{Manipulation of Bloch states}
Our shortcut method can also be employed to manipulate Bloch states~\cite{Deng,Xiong,Yue,Yang}, for instance to design $\pi/2$ or $\pi$ pulses between S and D bands constituting a pseudo-spin system. Unlike conventional Ramsey interferometer where selection rules can be used to prepare population in two states, the lattice band transition, similar to transition between vibration states in molecules~\cite{mole}, have no selection rules. For an arbitrary initial superposition between S and D band states, $|\psi_i\rangle=\sin\frac{\theta}{2}|S\rangle+e^{i\phi}\cos\frac{\theta}{2}|D\rangle$, a $\pi/2$ pulse $\hat{U}_{\pi/2}$ should transfer the initial states to the final states by
\begin{eqnarray}
\hat{U}_{\pi/2}|\psi_i\rangle=\frac{1}{\sqrt{2}}(\sin\frac{\theta}{2}+\cos\frac{\theta}{2})|S\rangle+\frac{1}{\sqrt{2}}e^{i\alpha}e^{i\phi}(\cos\frac{\theta}{2}-\sin\frac{\theta}{2})|D\rangle,\label{pi2}
\end{eqnarray}%
where $\alpha$ is the additional phase from $\hat{U}_{\pi/2}$.
If we choose $\alpha=\pi-\phi$, it is easy to prove that if $\hat{U}_{\pi/2}$ satisfies
\begin{eqnarray}
\hat{U}_\mathrm{\pi/2} |S\rangle =(|S\rangle+|D\rangle)/\sqrt{2}\ {\rm and} \ \hat{U}_\mathrm{\pi/2} |D\rangle = (-|S\rangle+|D\rangle)/\sqrt{2},\label{picon}
\end{eqnarray}
then $\hat{U}_{\pi/2}$ would satisfy Eq.(\ref{pi2}). By tuning the parameters of the pulse sequence, one can optimize the two fidelities $\zeta_{1}$=$|\langle \psi_{a_1} | \psi_{f_1}\rangle|^2$ and $\zeta_{2}$=$|\langle \psi_{a_2} | \psi_{f_2} \rangle|^2$ to their maximum, which is different from the above sequences with only one constraint, where $|\psi_{f_1}\rangle\equiv\hat{U}_\mathrm{\pi/2} |S\rangle$, $|\psi_{f_2}\rangle\equiv\hat{U}_\mathrm{\pi/2} |D\rangle$, $|\psi_{a_1}\rangle  \equiv  (|S\rangle+|D\rangle)/\sqrt{2}$ and $|\psi_{a_2}\rangle \equiv  (-|S\rangle+|D\rangle)/\sqrt{2}$.
For $V_0=10E_\mathrm{r}$ and a time sequence for $\pi/2$ pulse of $\{\textbf{56.2}, 28, \textbf{22.6}, 23.1\}(\mu$s) we find $\zeta_1=96.9\%$ and $\zeta_2=97.9\%$. The method for designing the $\pi$ pulse is similar to the case of the $\pi/2$ pulse.  Using a time sequence $\{\textbf{49.2}, 52.5, \textbf{22.1}, 26.4\}(\mu$s) we transfer atoms from $|S\rangle$ or $|D\rangle$ into the target states $|D\rangle$ or $-|S\rangle$ with the fidelity $98.5\%$ and $98.0\%$, respectively,. These time sequences have been employed to develop a matter wave Ramsey interferometer for motional quantum states exploiting the S- and D-bands of an OL ~\cite{Hu1}.

\section{Loading atoms into 2D and 3D optical lattice}\label{3D}

The above discussion has been focused on 1D optical lattices. We now show how these shortcut pulses can be extended to 2D or 3D OL, such as the square, triangular, hexagonal OLs~\cite{PRA50.5173,Becker,Struck,PRL108.045305}. Among these OLs, the square lattice is the simplest, where the total potential energy is the sum of potential energies in $\hat{k}_x$ and $\hat{k}_y$ directions, while this is not true for the triangular OL.

\subsection{2D square optical lattice}

The potential energy for 2D square OL can be divided into potential energies in the $\hat x$ and $\hat y$ directions if $||\vec{k}_x|-|\vec{k}_y||\ne 0$ or the electric field $ \vec{E}_1\perp\vec{E}_2$ in Eq.(\ref{e1}), in this case, the Hamiltonian of OL is given by
\begin{eqnarray} \label{e4-2d}
  \hat{H} =  - \frac{{{1}}}{{2m}}[\frac{{{\partial ^2}}}{{\partial {x^2}}}+ \frac{{{\partial ^2}}}{{\partial {y^2}}}] + \frac{1}{2}[V\left( x \right){\cos}\left( {{2k_x}x} \right)+ V\left( y \right){\cos}\left( {{2k_y}y} \right)],
\end{eqnarray}
and the wave functions can be separated in the form of $\psi  (\vec{r}) = \psi_x  (x)\psi_y (y)$.

\begin{figure}
 \includegraphics[width=1\textwidth]{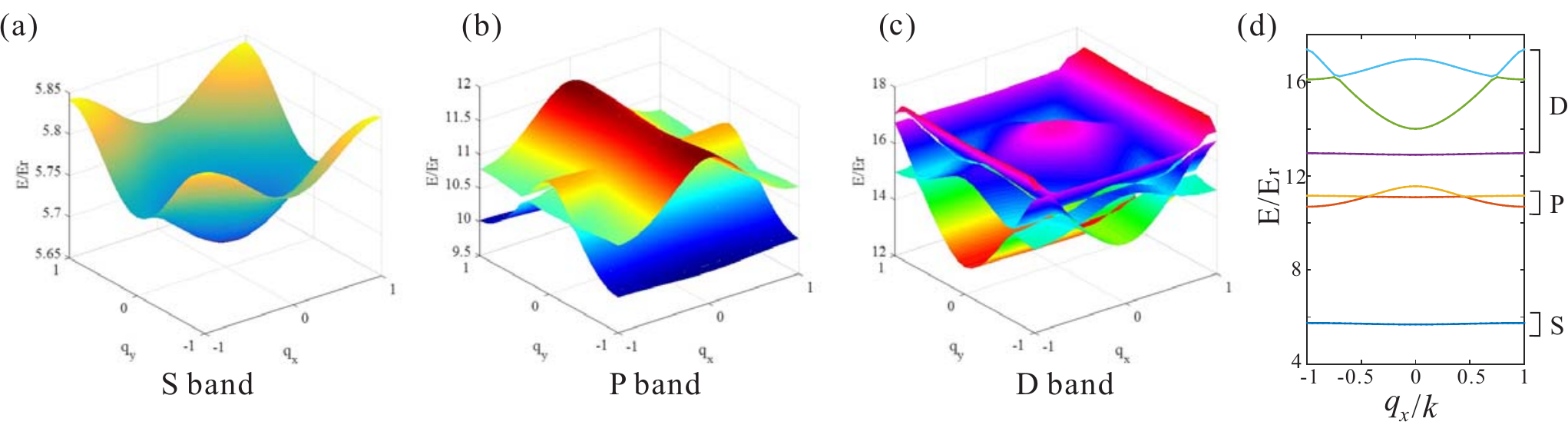}
  \caption{Schematic of Bloch bands in 2D square lattice: (a), (b) and (c) represent S-, P- and D-bands, respectively. (d) S-, P- and D-band energies along the $\hat{x}$ direction at $q_y=0$.}\label{f1_2d}
\end{figure}

\begin{figure}[b]
 \includegraphics[width=1\textwidth]{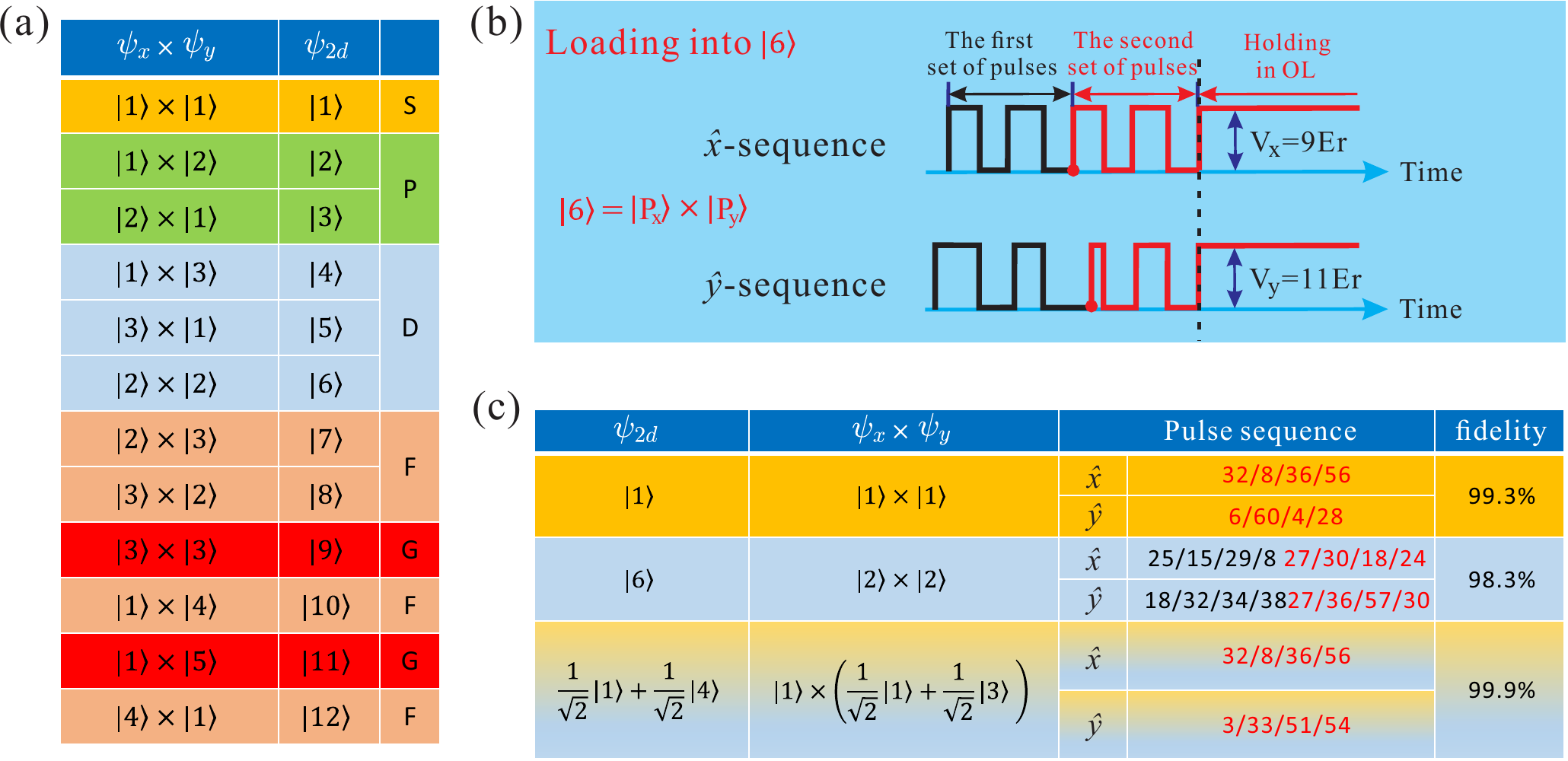}
  \caption{(a) The correspondence between 1D lattice and 2D square lattice band energy spectra with zero quasi-momentum. Bloch states of square lattice in the form of $\psi_x \times \psi_y$ in first column represent products of two 1D Bloch states. The Bloch state $\psi_{2d}$ in square lattice and its band are shown in the second and third columns, respectively. (b) The schematic diagram of the time sequences in the $\hat x$ and $\hat y$ directions corresponding to $\psi_{2d}=\left| 6 \right\rangle$. (c) The loading processes for three different target states and their corresponding calculated fidelities for $\psi_{2d}=\left| 1 \right\rangle$, $\left| 6 \right\rangle$ and
$\frac{1}{\sqrt{2}}\left| 1 \right\rangle+\frac{1}{\sqrt{2}}\left| 4 \right\rangle$. }\label{f2_2d}
\end{figure}

In Fig.~\ref{f1_2d} we draw the schematic of Bloch bands of the square lattice for $V_x=11E_r$ and $V_y=9E_r$. There is a difference between the lattice depths in the $\hat{x}$ and $\hat{y}$ directions in order to avoid energy degeneracy. The S-band is the ground band, and there are two P bands, P$_{x}$ and P$_{y}$, and three D bands D$_{x}$, D$_{y}$ and D$_{xy}$. Fig.~\ref{f1_2d}(d) shows the energy bands along the $\hat{k}_x$ direction at $q_y=0$. For Bloch states $\psi_{2d}$ in each band, we can arrange their eigenstates according to their eigenenergies, which are shown in the second column of Fig.~\ref{f2_2d}(a). The first state is S-band and the second and third states are $ \left| P_x \right\rangle$  and $\left| P_y \right\rangle  $ respectively. The 4th, 5th and 6th states are $ \left| D_x \right\rangle$, $ \left| D_y \right\rangle$ and $ \left| D_{xy} \right\rangle$ respectively. At the same time, the first column in Fig.~\ref{f2_2d}(a) displays the product forms of the corresponding two one-dimensional Bloch states.

For loading atoms into 2D square lattice, the evolution operator can be separated in the $\hat{x}$ and $\hat{y}$ directions $\widehat{U}\psi=\widehat{U}_x\psi_x\widehat{U}_y\psi_y$. If the target state in the square lattice $ |\psi_{a}\rangle =\sum\limits_{{i,j}}{\gamma_{ij}\left|n_x=i,n_y=j \right\rangle}$ can be written in the form of the product of two 1D states $|\psi_{a}\rangle=\sum\limits_{i}{\alpha_{i}\left|n_x=i\right\rangle}\otimes\sum\limits_{j}{\beta_{j}\left|n_y=j\right\rangle}$. Then
coefficients of these two forms should satisfy
\begin{equation}\label{coefficients}
\begin{bmatrix}
\alpha_{1}\\ \alpha_{2}\\ \vdots\\ \alpha_{m}
\end{bmatrix}
\begin{bmatrix}
\beta_{1}& \beta_{2}&\cdots& \beta_{m}
\end{bmatrix}
=
\begin{bmatrix}
\gamma_{11}&\gamma_{12}&\cdots&\gamma_{1m}\\ \gamma_{21}&\gamma_{22}&\cdots&\vdots\\ \vdots&\vdots&\ddots&\vdots\\ \cdots&\cdots&\cdots&\gamma_{mm}
\end{bmatrix}.
\end{equation}
We could use two separate 1D pulse sequences in the $\hat{x}$ and $\hat{y}$ directions to obtain target state with close to 100\% fidelity. If Eq.(\ref{coefficients}) does not hold, we can use numerical optimisation to obtain fidelities as high as possible. In Fig.~\ref{f2_2d}(b), we demonstrate this loading sequences for $\psi_{2d}=\left| 6 \right\rangle$, where the pulse sequences in two directions are independent but end at the same time, and the red part of the pulse sequence represents the laser with a phase shift that breaks the parity conservation. Fig.~\ref{f2_2d}(c) displays the calculated time sequences for three different target states and their fidelities that are very high.

\subsection{2D triangular optical lattice}

\begin{figure}[b]
\includegraphics[width=1\textwidth]{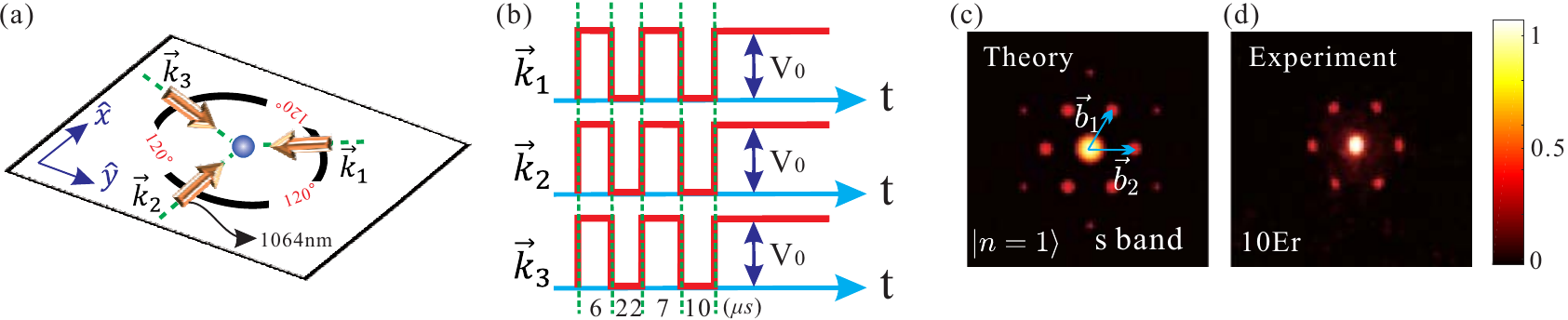}
  \caption{The demonstration for loading atoms into 2D triangle OL. (a) Sketch of 2D triangle OL: three travelling-wave lasers with an angle $120$ degree between each other intersect at one point on the $\hat{x}$-$\hat{y}$ plane. (b) The time sequence for loading atoms into S-band in triangular lattice, where the sequences for three beams $k_1$, $k_2$ and $k_3$ are the same. The corresponding population distribution in momentum space at $\vec{q}=0$ for state $\left| n=1 \right\rangle$ is shown as the calculated (c) and experimental results(d), respectively.
  }\label{f3_2d}
\end{figure}

For other 2D configurations, such as triangular OL constructed by three traveling-wave lasers with $|\vec{k}_1|=|\vec{k}_2|=|\vec{k}_3|=|\vec{k}|$ and $arg(\vec{k}_i,\vec{k}_j)=\pi/3$ ($i\ne j)$ (Fig.~\ref{f3_2d}(a)) we can't separate the variables in the $\hat x$ and $\hat y$ directions, and the Bloch states are written as
\begin{eqnarray}\label{e13-2d}
\left|{n,\vec{q}}\right\rangle =\sum_{\ell_1,\ell_2}c_{\ell_1,\ell_2}\left|{\ell_1 \vec{b}_1+\ell_2  \vec{b}_2+\vec{q}}\right\rangle.
\end{eqnarray}
where $\vec{b}_1=\sqrt{3}k\hat{x}$ and $\vec{b}_2=\sqrt{3}k(-\frac{1}{2}\hat{x},-\frac{\sqrt{3}}{2}\hat{y})$.  For the target state $\left| {\psi_a} \right\rangle =\sum_{n}\gamma_{n}\left|{n,\vec{q}}\right\rangle=\left|{1,0}\right\rangle$, we can impose the same time sequence on the three traveling beams as shown in Fig.~\ref{f3_2d}(b). The theoretical momentum distribution is shown in Fig.~\ref{f3_2d}(c) for $V_0=10E_r$. Using the time sequence $(t_{11},t_{12},t_{21},t_{22})=(\textbf{6},22,\textbf{7},10)\mu s$ we can reach the theoretical fidelity $\zeta=0.991$  for $V_0=10E_r$. The corresponding experimental image with NAS shown in Fig.~\ref{f3_2d}(d) is in agreement with the theoretical result.

\begin{figure}
 \includegraphics[width=0.95\textwidth]{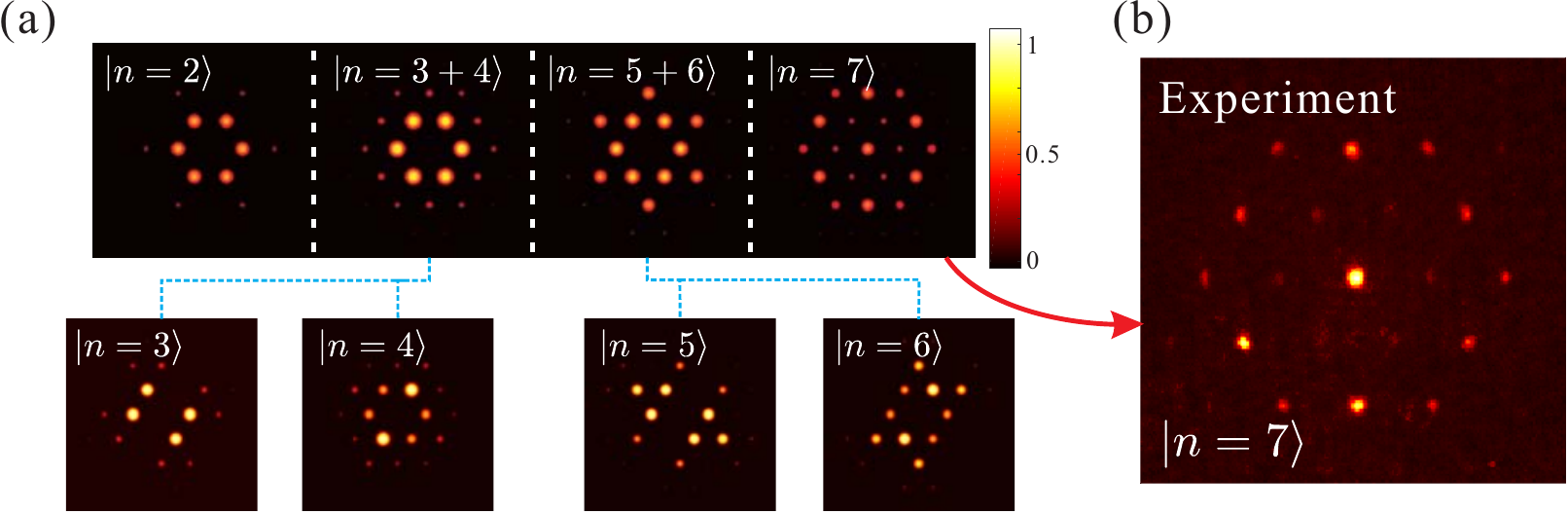}
  \caption{ (a) The calculated population distribution in momentum space for higher bands in 2D triangular lattice, where $\left| n=i \right\rangle$ (i=2,3...) represents the $i$th eigenstate. (b) Experimental population image for $|n=7\rangle$.}\label{f4_2d}
\end{figure}

The corresponding theoretical population distributions in momentum space for higher bands in a 2D triangular lattice are shown in Fig.~\ref{f4_2d}(a), where $\left| n=i \right\rangle$ (i=2,3...) represents the $i^{th}$ eigenstate with zero quasi-momentum, and $\left|n=3+4\right\rangle$ represents the superposition of degenerate states, $\left| n=3 \right\rangle$ and $\left| n=4 \right\rangle$. For instance, if we choose the target state $\left| {\psi_a} \right\rangle =\left|{7,0}\right\rangle$, we can get $\zeta=0.92$ using time sequence $(t_{11},t_{12},t_{21},t_{22})=(\textbf{22.1},37.9,\textbf{79.9},35.6)\mu s$ for the lattice depth $V_0=10E_r$. The experimental results are shown in Fig.~\ref{f4_2d}(b). To load atoms into other excited states, more complicated pulse sequences with different phases are required.

\subsection{3D optical lattice}

\begin{figure}[t]
 \begin{center}
 \includegraphics[width=1\textwidth]{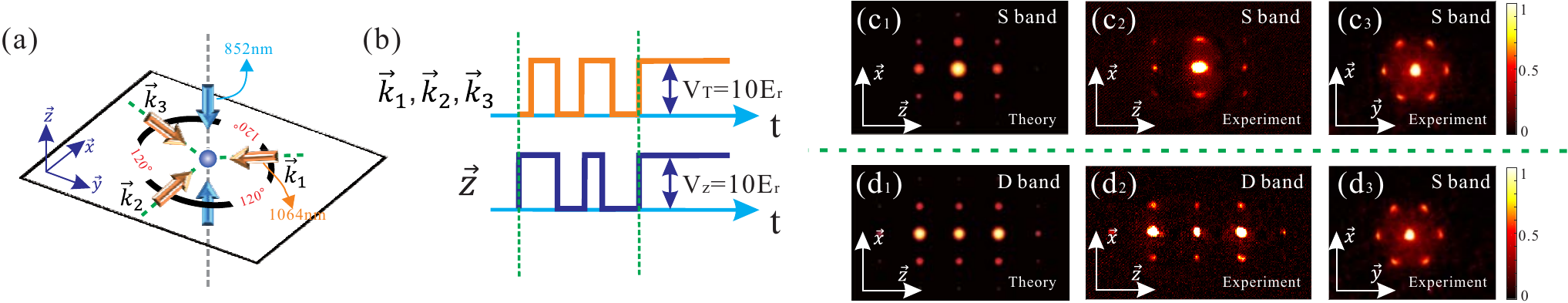}
  \end{center}
  \caption{ (a) Sketch of the 3D triangular OL: an optical standing waves with $852nm$ parallel to the $z$-axis, and three travelling waves with $1064 nm$ that intersect in the $\hat{x}$-$\hat{y}$ plane. (b) Time sequences for the $\hat{x}$-$\hat{y}$ plane and the $\hat{z}$ direction are different. (c) The calculated and experimental populations for S-band in 3D lattice, (c1) the calculated momentum distributions in the $\hat{x}$-$\hat{z}$ plane, and (c2) and (c3) measured absorption images by NAS in the $\hat{y}$ direction and the $\hat{z}$ direction, respectively. (d) The case of S-band in the $\hat{x}$-$\hat{y}$ plane and D-band in the $\hat z$ direction. (d1-d3) is similar to (c1-c3).
  }\label{f5_2d}
\end{figure}

For the simplest 3D cubic OL, the wave functions can be separated by variables in the $\hat{x}$, $\hat{y}$ and $\hat{z}$ directions, and we can also load BEC to arbitrary target states. A more complicated 3D lattice composed of a 2D triangular lattice in the $\hat{x}$-$\hat{y}$ plane with $\lambda=1064nm$ and a 1D lattice in the $\hat{z}$ direction with $\lambda=852nm$ is shown in Fig.~\ref{f5_2d}(a). For this OL we can combine the 1D sequence and 2D sequence. Fig.~\ref{f5_2d}(b) shows the different time sequences on the $\hat{x}$-$\hat{y}$ plane and the $\hat{z}$ direction. The atoms can be transferred from the harmonic trap into the S-band of the OL in the $\hat{x}$-$\hat{y}$ plane and $\hat{z}$ direction, as shown in Fig.~\ref{f5_2d}(c), or S-band in the $\hat{x}$-$\hat{y}$ plane and D-band in the $\vec{z}$ direction as shown in Fig.~\ref{f5_2d}(d). The time sequences in Fig.~\ref{f5_2d}(c) are $(\textbf{6},22,\textbf{7},10)\mu s$ in the $\hat{x}$-$\hat{y}$ plane and $(\textbf{24.5},28.8,\textbf{8.1},2.2)\mu s$ in the $\hat{z}$ direction. The sequences in Fig.~\ref{f5_2d}(d) are $(\textbf{6},22,\textbf{7},10)\mu s$ in the $\hat{x}$-$\hat{y}$ plane and $(\textbf{5.5}, 21.0, \textbf{13.0}, 6.1) \mu s$ in the $\hat{z}$ direction. The experimental results are in agreement with the theoretical calculations.

\section{Conclusions}\label{Conclusions}

In summary, we present a method for effective preparation of a BEC in different bands of an optical lattice within a few tens of microseconds. This shortcut stems from nonholonomic coherent control, composed by pulse sequences which are imposed on the system before the OL switches on and fully optimised for high fidelity and robustness.  With our approach, the BEC can be prepared in either pure Bloch states or superposition of states of different bands.
Furthermore we show this shortcut can also be successfully applied for 2D and 3D OLs. The experimental results are well described by the theoretical calculations . Because the duration of pulses is short enough, the atom-atom interaction can be neglected during the design of pulse sequences, and the numerical results show that the interaction leads to a change of fidelity less than $1\%$ in our designed time sequence. This efficient shortcut not only provides applications in controllable quantum systems and quantum information processing, but also is helpful for the study of orbital optical lattices, simulation of systems in condensed matter physics, and the precise measurements.

\ack{ We thank L. Yin, P. Zhang, B. Wu, G. J. Dong, H. W. Xiong and X. Chen for useful discussions. This work is supported by the National Key Research and Development Program of China (Grant
No. 2016YFA0301501), and the NSFC (Grants No.11334001, No.61475007 and No. 61727819).
JS acknowledges support by the European Research Council, ERC-AdG, Quantum Relax. }


\newpage
\section*{References}
\begin{thebibliography}{<59>}

\bibitem{BlochRev} I. Bloch, J. Dalibard and W. Zwerger 2008  {\it Rev. Mod. Phys.} \textbf{80} 885.
\bibitem{Dervianko} A.  Derevianko and H.  Katori 2011 {\it Rev. Mod. Phys.} \textbf{83} 331.
\bibitem{Cronin} A. D. Cronin, J. Schmiedmayer and D. E. Pritchard 2009 {\it Rev. Mod. Phys.} {\bf 81} 1051.
\bibitem{book} C. Cohen-Tannoudji and D. Gu\'ery-Odelin 2011 \textit{Advances in atonic physics: An overview}, {\it World Scientific}.

\bibitem{reviewSTA13} E. Torrontegui, S. Ibanez, S. Martinez-Garaot, M. Modugno, A. del Campo, D. Gu\'ery-Odelin, A. Ruschhaupt, X. Chen and J. G. Muga 2013 {\it Adv. Atom. Mol. Opt. Phys.} \textbf{62} 117.
\bibitem{Chen} X. Chen, A. Ruschhaupt, S. Schmidt, A. del Campo, D. Gu\'ery-Odelin and J. G. Muga 2010 {\it Phys. Rev. Lett.} \textbf{104} 063002.
\bibitem{Schaff-2} J. F. Schaff, X. L. Song, P. Capuzzi, P. Vignolo and G. Labeyrie 2011 {\it Europhys. Lett.} \textbf{93}, 23001.
\bibitem{Oliver} M. G. Bason, M. Viteau, N. Malossi, P. Huillery, E. Arimondo, D. Ciampini, R. Fazio, V. Giovannetti, R. Mannella and O. Morsch 2012 {\it Nat. Phys.} \textbf{8} 147.
\bibitem{Sofia} S. Mart\'{i}nez-Garaot, E. Torrontegui, X. Chen, M. Modugno, D. Gu\'{e}ery-Odelin, S. Y. Tseng and J. G. Muga 2013 {\it Phys. Rev. Lett.} \textbf{111} 213001.
\bibitem{Campo} S. Masuda, K. Nakamura and A. del Campo 2014 {\it Phys. Rev. Lett.} \textbf{113} 063003.
\bibitem{JorgeSciRep} W. Rohringer, D. Fischer, F. Steiner, I. E. Mazets, J. Schmiedmayer and  M. Trupke 2015 {\it Sci. Rep.} \textbf{5} 9820.
\bibitem{Masuda} S. Masuda and S. A. Rice 2015 {\it J. Phys. Chem. B} \textbf{119} 11088.
\bibitem{Strungari} D. J. Papoular and S. Stringari 2015 {\it Phys. Rev. Lett.} \textbf{115} 025302.

\bibitem{Rabitz} A. Bulatov, B. Vugmeister, A. Burin and H. Rabitz 1999 {\it Phys. Rev. A} \textbf{60} 4875.
\bibitem{Jorge}  R. Bucker, T. Berrada, S. van Frank, J.-F. Schaff, T. Schumm, J. Schmiedmayer, G. J\"{a}ger, J. Grond and U. Hohenester 2013 {\it J. Phys. B: At. Mol. Opt. Phys.} \textbf{46} 104012.

\bibitem{Levitt}  M.H. Levitt 1986 {\it Prog. Nucl. Magn. Reson. Spectrosc.} \textbf{18} 61.

\bibitem{Schleier} M. H. Schleier-Smith, I. D. Leroux and V. Vuleti\'{c} 2010 {\it Phys. Rev. Lett.} \textbf{104} 073604.
\bibitem{Butts} D. L. Butts, K. Kotru, J. M. Kinast, A. M. Radojevic, B. P. Timmons and R. E. Stoner 2013 {\it J. Opt. Soc. Am. B} \textbf{30} 922.
\bibitem{Lee} J. H. Lee, E. Montano, I.H. Deutsch and P. S. Jessen 2013 {\it Nat. Commun.} \textbf{4} 2027.
\bibitem{Takahashi} S. Taie, H. Ozawa, T. Ichinose, T. Nishio, S. Nakajima and  Y. Takahashi 2015 {\it Sci. Adv.} \textbf{1} e1500854.
\bibitem{Scarola} V. W. Scarola and S. Das Sarma 2005 {\it Phys. Rev. Lett.} {\bf 95}, 033003.



\bibitem{Wu}  C. Wu, W. V. Liu, J. Moore and S. Das Sarma 2006 {\it Phys. Rev. Lett.} {\bf 97} 190406.

\bibitem{Lewenstein}  M. Lewenstein and W. V. Liu 2011 {\it Nat. Phys.} {\bf 7} 101.


\bibitem{Muller} T. M\"uller, S. F\"olling, A. Widera and I. Bloch 2007 {\it Phys. Rev. Lett.} {\bf 99} 200405.

\bibitem{Browaeys} A. Browaeys, H. H\"affner, C. McKenzie, S. L. Rolston, K. Helmerson and W. D. Phillips 2005 {\it Phys. Rev. A} {\bf 72} 053605.
\bibitem{Wirth1} G. Wirth, Molschlager and A. Hemmerich 2011 {\it Nature Phys.} {\bf 7} 147.
\bibitem{M} M. \"{O}lschl\"{a}ger, G. Wirth and A. Hemmerich 2011 {\it Phys. Rev. Lett.} {\bf 106} 015302.

\bibitem{Lloyd} S. Lloyd 1995 {\it Phys. Rev. Lett.} {\bf
75} 346.

\bibitem{Harel} G. Harel and V. M. Akulin 1999 {\it Phys. Rev. Lett.} {\bf
82} 1.

\bibitem{JPB02} J. H. Denschlag, J. E. Simsarian, H. Haffner, C. McKenzie, A. Browacys, D. Cho, K. Helmerson, S. L. Rolston and W. D. Phillips 2002 {\it J. Phys. B: At. Mol. Opt. Phys.} {\bf 35} 3095.

\bibitem{TN} T.No\"bauer, A. Angerer, B. Bartels, M. Trupke, S. Rotter, J. Schmiedmayer, F. Mintert and J. Majer 2015 {\it Phys. Rev. Lett.} {\bf 115} 190801.

\bibitem{EsslingerMap1} M. K\"{o}hl, H. Moritz, T. St\"{o}ferle, K. G\"{u}nter and T. Esslinger 2005 {\it Phys. Rev. Lett.} {\bf 94} 080403.
\bibitem{SpreeuwMap} A. Kastberg, W. D. Phillips, S. L. Rolston, R. J. C. Spreeuw and P. S. Jessen 1995 {\it Phys. Rev. Lett.} {\bf 74} 1542.
\bibitem{EsslingerMap2} M. Greiner,  I. Bloch, O. Mandel, T. W. H\"{a}nsch and T. Esslinger 2001 {\it Phys. Rev. Lett.} {\bf 87} 160405.


\bibitem{Liu} X. X. Liu, X. J. Zhou, W. Xiong, T. Vogt and X. Z. Chen 2011 {\it Phys. Rev. A} {\bf 83} 063402.




\bibitem{Zhai} Y. Y. Zhai, X. G. Yue, Y. J. Wu, X. Z. Chen, P. Zhang and X. J. Zhou 2013 {\it Phys. Rev. A} {\bf 87} 063638.

\bibitem{Wang} Z. K. Wang, B. G. Yang, D. Hu, X. Z. Chen, H. W. Xiong, B. Wu and X. J. Zhou 2016 {\it Phys. Rev. A} {\bf 94} 033624.
\bibitem{Bloch2} T. M\"{u}ller, Simon F\"{o}lling, A. Widera and I. Bloch 2007 {\it Phys. Rev. Lett.} {\bf99} 200405.
\bibitem{speed} A. Browaeys, H. H\"{a}ffner, C. McKenzie, S. L. Rolston, K. Helmerson and W. D. Phillips 2005 {\it Phys. Rev. A} {\bf72} 053605.
\bibitem{NP} P. S. Panahi, D. S. L\"u\"hmann, J. Struck, P. Windpassinger and K. Sengstock 2012 {\it Nat. Phys.} {\bf8} 71.

\bibitem{shaken} C. V. Parker, L. Ha and C. Chin 2013 {\it Nat. Phys.} {\bf9} 769.

\bibitem{GW} G. Wirth, M. \"Olschl\"ager and A. Hemmerich 2011 {\it Nat. Phys.} {\bf7} 147.
\bibitem{Hu} D. Hu, L. X. Niu, B. G. Yang, X. Z. Chen, B. Wu, H. W. Xiong and X. J. Zhou 2015 {\it Phys. Rev. A} {\bf 92} 043614.

\bibitem{Niu1} L. X. Niu, P. J. Tang, B. G. Yang, X.Z.Chen, B. Wu and X. J. Zhou 2016 {\it Phys. Rev. A} {\bf 94} 063603.
\bibitem{Hu1} D. Hu, L. X. Niu, S. J. Jin, X. Z. Chen, G. J. Dong, J. Schmiedmayer and X. J. Zhou 2017, arXiv:2110028.
\bibitem{Deng} L. Deng, E. W. Hagley, J. Denschlag, J. E. Simsarian, M. Edwards, C. W. Clark, K. Helmerson, S. L. Rolston and W. D. Phillips 1999 {\it Phys. Rev. Lett.} {\bf 83} 5407.
\bibitem{Xiong} W. Xiong, X. G. Yue, Z. K. Wang, X. J. Zhou and X. Z. Chen 2011 {\it Phys. Rev. A} {\bf 84} 043616.
\bibitem{Yue} X. G. Yue, Y. Y. Zhai, Z. K. Wang, H. W. Xiong, X. Z. Chen and X. J. Zhou 2013 {\it Phys. Rev. A} {\bf 88} 013603.
\bibitem{Yang} B. G Yang, S. J. Jin, X. Y. Dong, Z. Liu, L. Yin and X. J. Zhou 2016 {\it Phys. Rev. A} {\bf 94} 043607.
\bibitem{mole} K. Kneipp, Y. Wang, H. Kneipp, I. Itzkan, R. R. Dasari and M. S. Feld 1996 {\it Phys. Rev. Lett.} {\bf 76} 2444.

\bibitem{PRA50.5173} K. I. Petsas, A. B. Coates and G. Grynberg 1994 {\it Phys. Rev. A} {\bf 50} 5173.



\bibitem{Becker} C. Becker, P. Soltan-Panahi, J. Kronj\"ager, S. D\"orscher, K. Bongs and K. Sengstock 2010 {\it New J. Phys.} {\bf 12} 065025.

\bibitem{Struck} J. Struck, C. \"Olschl\"ager, R. Le Targat, P. Soltan-Panahi, A. Eckardt, M. Lewenstein and K. Sengstock 2011 {\it Science} {\bf 333} 996.

\bibitem{PRL108.045305} G. B. Jo, J. Guzman, C. K. Thomas, P. Hosur, A. Vishwanath and D. M. Stamper-Kurn 2012 {\it Phys. Rev. Lett.} {\bf 108} 045305.





\end{thebibliography}
\end{document}